\documentclass[twocolumn,aps,prc,superscriptaddress,nofootinbib,showpacs,floatfix,longbibliography]{revtex4-1}
\usepackage{url}
\usepackage{cancel}
\usepackage[colorlinks,linkcolor=blue,citecolor=blue,filecolor=black,urlcolor=blue]{hyperref}
\usepackage{epsfig,graphics}
\usepackage{graphicx}% Include figure files
\usepackage{dcolumn}% Align table columns on decimal point
\usepackage{bm}% bold math
\usepackage{amssymb}
\usepackage{amsmath}
\usepackage{multirow}
\usepackage{float}
\usepackage{harpoon}
\usepackage{MnSymbol}
\usepackage{appendix}
\usepackage{color}
\usepackage{hyperref}
\usepackage{cleveref}

\newcommand{\Yphi}{$Y(\Delta\phi)$}
\newcommand{\YphiTempl}{Y(\Delta\phi)^{\mathrm{templ}}}
\newcommand{\YphiRidge}{Y(\Delta\phi)^{\mathrm{ridge}}}

\newcommand{\Yphipp}{Y(\Delta\phi)^{pp}}

\newcommand{\dphi}{\Delta\phi}

\newcommand{\TopE}{$\sqrt{s_{_{\mathrm{NN}}}}=200$~GeV}

\newcommand{\snn}{\mbox{$\sqrt{s_{\mathrm{NN}}}$}}
\newcommand{\npart}{\mbox{$N_{\mathrm{part}}$}}

\newcommand{\pT} {p_{\mathrm{T}}}
\newcommand{\pTt} {p_{\mathrm{T}}^{\mathrm t}}
\newcommand{\pTa} {p_{\mathrm{T}}^{\mathrm a}}

\newcommand{\lr}[1]{\left\langle #1\right\rangle}

\newcommand{\Dphi}{\mbox{$\Delta \phi$}}
\newcommand{\Deta}{\mbox{$\Delta \eta$}}
\newcommand{\nch}{\lr{N_{\mathrm{ch}}}}

\newcommand{\heau}{$^{3}$He$+$Au}
\newcommand{\dau}{$d+\rm{Au}$}
\newcommand{\pp}{$p+p$}
\newcommand{\pau}{$p+\rm{Au}$}
\usepackage{lineno}
\usepackage{CJK}
\begin{document}
\title{Measurement of flow coefficients in high-multiplicity $p$+Au, $d$+Au and $^{3}$He$+$Au collisions at $\sqrt{s_{_{\mathrm{NN}}}}$=200~GeV }

\affiliation{Abilene Christian University, Abilene, Texas   79699}
\affiliation{Alikhanov Institute for Theoretical and Experimental Physics NRC "Kurchatov Institute", Moscow 117218}
\affiliation{Argonne National Laboratory, Argonne, Illinois 60439}
\affiliation{American University in Cairo, New Cairo 11835, Egypt}
\affiliation{Ball State University, Muncie, Indiana, 47306}
\affiliation{Brookhaven National Laboratory, Upton, New York 11973}
\affiliation{University of Calabria \& INFN-Cosenza, Rende 87036, Italy}
\affiliation{University of California, Berkeley, California 94720}
\affiliation{University of California, Davis, California 95616}
\affiliation{University of California, Los Angeles, California 90095}
\affiliation{University of California, Riverside, California 92521}
\affiliation{Central China Normal University, Wuhan, Hubei 430079 }
\affiliation{University of Illinois at Chicago, Chicago, Illinois 60607}
\affiliation{Chongqing University, Chongqing, 401331}
\affiliation{Creighton University, Omaha, Nebraska 68178}
\affiliation{Czech Technical University in Prague, FNSPE, Prague 115 19, Czech Republic}
\affiliation{National Institute of Technology Durgapur, Durgapur - 713209, India}
\affiliation{ELTE E\"otv\"os Lor\'and University, Budapest, Hungary H-1117}
\affiliation{Frankfurt Institute for Advanced Studies FIAS, Frankfurt 60438, Germany}
\affiliation{Fudan University, Shanghai, 200433 }
\affiliation{Guangxi Normal University, Guilin, 541004}
\affiliation{University of Heidelberg, Heidelberg 69120, Germany }
\affiliation{University of Houston, Houston, Texas 77204}
\affiliation{Huzhou University, Huzhou, Zhejiang  313000}
\affiliation{Indian Institute of Science Education and Research (IISER), Berhampur 760010 , India}
\affiliation{Indian Institute of Science Education and Research (IISER) Tirupati, Tirupati 517507, India}
\affiliation{Indian Institute Technology, Patna, Bihar 801106, India}
\affiliation{Indiana University, Bloomington, Indiana 47408}
\affiliation{Institute of Modern Physics, Chinese Academy of Sciences, Lanzhou, Gansu 730000 }
\affiliation{University of Jammu, Jammu 180001, India}
\affiliation{Joint Institute for Nuclear Research, Dubna 141 980}
\affiliation{Kent State University, Kent, Ohio 44242}
\affiliation{University of Kentucky, Lexington, Kentucky 40506-0055}
\affiliation{Lawrence Berkeley National Laboratory, Berkeley, California 94720}
\affiliation{Lehigh University, Bethlehem, Pennsylvania 18015}
\affiliation{Max-Planck-Institut f\"ur Physik, Munich 80805, Germany}
\affiliation{Michigan State University, East Lansing, Michigan 48824}
\affiliation{National Research Nuclear University MEPhI, Moscow 115409}
\affiliation{National Institute of Science Education and Research, HBNI, Jatni 752050, India}
\affiliation{National Cheng Kung University, Tainan 70101 }
\affiliation{The Ohio State University, Columbus, Ohio 43210}
\affiliation{Panjab University, Chandigarh 160014, India}
\affiliation{NRC "Kurchatov Institute", Institute of High Energy Physics, Protvino 142281}
\affiliation{Purdue University, West Lafayette, Indiana 47907}
\affiliation{Rice University, Houston, Texas 77251}
\affiliation{Rutgers University, Piscataway, New Jersey 08854}
\affiliation{University of Science and Technology of China, Hefei, Anhui 230026}
\affiliation{South China Normal University, Guangzhou, Guangdong 510631}
\affiliation{Sejong University, Seoul, 05006, South Korea}
\affiliation{Shandong University, Qingdao, Shandong 266237}
\affiliation{Shanghai Institute of Applied Physics, Chinese Academy of Sciences, Shanghai 201800}
\affiliation{Southern Connecticut State University, New Haven, Connecticut 06515}
\affiliation{State University of New York, Stony Brook, New York 11794}
\affiliation{Instituto de Alta Investigaci\'on, Universidad de Tarapac\'a, Arica 1000000, Chile}
\affiliation{Temple University, Philadelphia, Pennsylvania 19122}
\affiliation{Texas A\&M University, College Station, Texas 77843}
\affiliation{University of Texas, Austin, Texas 78712}
\affiliation{Tsinghua University, Beijing 100084}
\affiliation{University of Tsukuba, Tsukuba, Ibaraki 305-8571, Japan}
\affiliation{University of Chinese Academy of Sciences, Beijing, 101408}
\affiliation{Valparaiso University, Valparaiso, Indiana 46383}
\affiliation{Variable Energy Cyclotron Centre, Kolkata 700064, India}
\affiliation{Wayne State University, Detroit, Michigan 48201}
\affiliation{Wuhan University of Science and Technology, Wuhan, Hubei 430065}
\affiliation{Yale University, New Haven, Connecticut 06520}

\author{M.~I.~Abdulhamid}\affiliation{American University in Cairo, New Cairo 11835, Egypt}
\author{B.~E.~Aboona}\affiliation{Texas A\&M University, College Station, Texas 77843}
\author{J.~Adam}\affiliation{Czech Technical University in Prague, FNSPE, Prague 115 19, Czech Republic}
\author{J.~R.~Adams}\affiliation{The Ohio State University, Columbus, Ohio 43210}
\author{G.~Agakishiev}\affiliation{Joint Institute for Nuclear Research, Dubna 141 980}
\author{I.~Aggarwal}\affiliation{Panjab University, Chandigarh 160014, India}
\author{M.~M.~Aggarwal}\affiliation{Panjab University, Chandigarh 160014, India}
\author{Z.~Ahammed}\affiliation{Variable Energy Cyclotron Centre, Kolkata 700064, India}
\author{A.~Aitbaev}\affiliation{Joint Institute for Nuclear Research, Dubna 141 980}
\author{I.~Alekseev}\affiliation{Alikhanov Institute for Theoretical and Experimental Physics NRC "Kurchatov Institute", Moscow 117218}\affiliation{National Research Nuclear University MEPhI, Moscow 115409}
% ...

\collaboration{STAR Collaboration}\noaffiliation

\begin{abstract}
Flow coefficients ($v_2$ and $v_3$) are measured in high-multiplicity $p$+Au, $d$+Au, and $^{3}$He$+$Au collisions at a center-of-mass energy of $\sqrt{s_{_{\mathrm{NN}}}}$ = 200 GeV using the STAR detector. The measurements utilize two-particle correlations with a pseudorapidity requirement of $|\eta| <$ 0.9 and a pair gap of $|\Delta\eta|>1.0$. The primary focus is on analysis methods, particularly the subtraction of non-flow contributions. Four established non-flow subtraction methods are applied to determine $v_n$, validated using the HIJING event generator. $v_n$ values are compared across the three collision systems at similar multiplicities; this comparison cancels the final state effects and isolates the impact of initial geometry. While $v_2$ values show differences among these collision systems, $v_3$ values are largely similar, consistent with expectations of subnucleon fluctuations in the initial geometry. The ordering of $v_n$ differs quantitatively from previous measurements using two-particle correlations with a larger rapidity gap, which, according to model calculations, can be partially attributed to the effects of longitudinal flow decorrelations. The prospects for future measurements to improve our understanding of flow decorrelation and subnucleonic fluctuations are also discussed.
\end{abstract}
\maketitle

\section{Introduction}\label{sec:1}
High-energy collisions of heavy nuclei, such as gold at RHIC and lead at the LHC, produce a hot and dense state of matter composed of strongly interacting quarks and gluons, known as the Quark-Gluon Plasma (QGP)~\cite{Busza:2018rrf}. This QGP undergoes rapid transverse expansion, converting initial spatial nonuniformities into significant anisotropic particle flow in transverse momentum ($\pT$). This flow can be accurately described by viscous relativistic hydrodynamic equations with extremely low viscosity~\cite{Gale:2013da,Heinz:2013th}. Therefore, the QGP is often referred to as a nearly inviscid liquid or ``perfect fluid''.

Experimentally, the anisotropic flow manifests as a harmonic modulation of particle distribution in the azimuthal angle $\phi$ for each event, described by:
\begin{align}\label{eq:1}
 \frac{dN}{d\phi}\propto 1+2\sum_{n=1}^{\infty}v_{n} (\pT)\cos(n(\phi-\Psi_{n}))\;.
\end{align}
Here, $v_n$ and $\Psi_n$ denote the magnitude and orientation of the $n^{\mathrm{th}}$-order harmonic flow, represented by the flow vector $V_n \equiv v_n e^{in\Psi_n}$. The most significant flow coefficients are the elliptic flow $v_2$ and the triangular flow $v_3$. However, measuring the event-wise distribution described by Eq.~\ref{eq:1} is limited by the finite number of particles produced in each event, so flow coefficients are obtained via a two-particle azimuthal correlation method:
\begin{align}\label{eq:2}
\frac{dN_{\mathrm{pairs}}}{d\dphi}\propto 1+2\sum_{n=1}^{\infty}c_{n} (\pTt,\pTa)\cos (n\dphi)\;,
\end{align}
where $c_n(\pT^{\mathrm t},\pT^{\mathrm a})=v_n(\pT^{\mathrm t})v_n(\pT^{\mathrm a})$, assuming a factorization behavior for $v_n$ extracted from two distinct $\pT$ ranges~\cite{ATLAS:2012at}. To mitigate short-range ``non-flow'' correlations from sources such as jet fragmentation and resonance decays, a pseudorapidity gap is employed between particles labeled as ``t'' (trigger) and ``a'' (associated).

Key questions include the minimum system size at which ``perfect fluid'' behavior can be observed and whether QGPs created with various sizes exhibit consistent properties. To address these questions, a series of measurements have been conducted in small systems ranging from $p$+$p$~\cite{CMS:2010ifv,ATLAS:2015hzw,CMS:2016fnw} to $p$+A~\cite{CMS:2012qk,ALICE:2012eyl,Aad:2012gla,PHENIX:2013ktj,PHENIX:2014fnc,PHENIX:2022nht,STAR:2022pfn}, and $\gamma$+A collisions~\cite{ATLAS:2021jhn}. These measurements revealed significant anisotropic flow in all these systems. The scientific community debates whether the observed flow originates from final-state effects (FS) due to the collective response to initial geometrical fluctuations, or from initial-state effects (IS), such as intrinsic momentum correlations within the nuclear wavefunction at high energies~\cite{Dusling:2015gta,Schenke:2021mxx}. IS models primarily rooted in gluon saturation physics exhibit short-range features in pseudorapidity $\eta$ and fail to reproduce detailed $\pT$ dependence~\cite{Schenke:2015aqa,Mace:2018vwq} and multi-particle correlations~\cite{Mace:2019rtt}. Consequently, the consensus leans towards the FS interpretation of collective flow in small systems.

The FS perspective does not necessarily imply hydrodynamics and the presence of a perfect fluid. Discussions revolve around whether the medium in these systems can be characterized as a perfect fluid with well-defined transport properties or as partons undergoing a few scatterings without achieving hydrodynamic or thermal equilibrium~\cite{He:2015hfa,Kurkela:2018qeb,Romatschke:2018wgi,Kurkela:2021ctp}. Some hydrodynamics models further introduce a ``pre-flow'' phase, where partons undergo free streaming before hydrodynamic evolution~\cite{Weller:2017tsr}. Despite differences, all these models assume that harmonic flow originates from initial spatial anisotropies, characterized by eccentricity vectors ${\mathcal{E}}_n=\varepsilon_n {e}^{{\textrm i}n\Phi_n}$.

In large collision systems, model calculations establish a linear relationship between flow and eccentricity for elliptic and triangular flow~\cite{Gardim:2011xv,Niemi:2012aj}:
\begin{align}\label{eq:3}
v_n=k_n \varepsilon_n, \; n=2,3\;.
\end{align}
Response coefficients $k_n$ encompass all final state effects and remain constant for events with similar particle multiplicities. However, Eq.~\ref{eq:3}'s validity and hydrodynamic interpretation are less established in small systems. Non-hydrodynamic approaches mentioned earlier can significantly alter the response coefficients $k_n$ and their $\pT$ dependencies. Distinguishing non-equilibrium transport from hydrodynamics is crucial to confirming perfect fluid behavior in small systems.

\begin{table}[tbh]
\caption{The values of $\varepsilon_{2}$ and $\varepsilon_{3}$ in central collisions (requiring either impact parameter $b<2$~fm or 0--5\% centrality), obtained from Glauber models~\cite{PHOBOSGlauber} including nucleon~\cite{Nagle:2013lja,PHENIX:2018lia,STAR:2022pfn} or subnucleon fluctuations~\cite{Welsh:2016siu}. They are defined either as simple average, $\lr{\varepsilon_{n}}$~\cite{Nagle:2013lja,PHENIX:2018lia}, or the root-mean-square values, $\sqrt{\lr{\varepsilon_{n}^2}}$, which take into account event-by-event fluctuations. The values have negligible statistical uncertainties. The values in 0--2\% or 0--10\% centralities are not shown, but they are nearly identical to those quoted for 0--5\%.}
\begin{ruledtabular} \begin{tabular}{ccccc}
&&$^{3}$He+Au&$d$+Au&$p$+Au \\\hline
Nucleon &\multirow{1}{*}{$\lr{\varepsilon_{2}}$}&\multirow{1}{*}{0.50}&\multirow{1}{*}{0.54}&\multirow{1}{*}{0.23}\\
Glauber~\cite{Nagle:2013lja,PHENIX:2018lia} &&&&\\
$b<2$~fm&\multirow{1}{*}{$\lr{\varepsilon_{3}}$}&\multirow{1}{*}{0.28}&\multirow{1}{*}{0.18}&\multirow{1}{*}{0.16}\\\hline

&\multirow{2}{*}{$\lr{\varepsilon_{2}}$}&\multirow{2}{*}{0.49}&\multirow{2}{*}{0.55}&\multirow{2}{*}{0.25}\\
&&&&\\
Nucleon&\multirow{2}{*}{$\lr{\varepsilon_{3}}$}&\multirow{2}{*}{0.29}&\multirow{2}{*}{0.23}&\multirow{2}{*}{0.20}\\
Glauber~\cite{PHOBOSGlauber,STAR:2022pfn}&&&&\\
0--5\% centrality&\multirow{2}{*}{$\sqrt{\lr{\varepsilon_{2}^2}}$}&\multirow{2}{*}{0.53}&\multirow{2}{*}{0.59}&\multirow{2}{*}{0.28}\\
&&&&\\
&\multirow{2}{*}{$\sqrt{\lr{\varepsilon_{3}^2}}$}&\multirow{2}{*}{0.33}&\multirow{2}{*}{0.28}&\multirow{2}{*}{0.23}\\
&&&&\\\hline

Subnucleon&\multirow{2}{*}{$\sqrt{\lr{\varepsilon_{2}^2}}$}&\multirow{2}{*}{0.54}&\multirow{2}{*}{0.55}&\multirow{2}{*}{0.41}\\
Glauber~\cite{Welsh:2016siu}&&&&\\
0--5\% centrality&\multirow{2}{*}{$\sqrt{\lr{\varepsilon_{3}^2}}$}&\multirow{2}{*}{0.38}&\multirow{2}{*}{0.35}&\multirow{2}{*}{0.34}\\
&&&&\\
\end{tabular} \end{ruledtabular}
\label{tab:1}
\end{table}

\begin{figure}[h]
\includegraphics[width=1.0\linewidth]{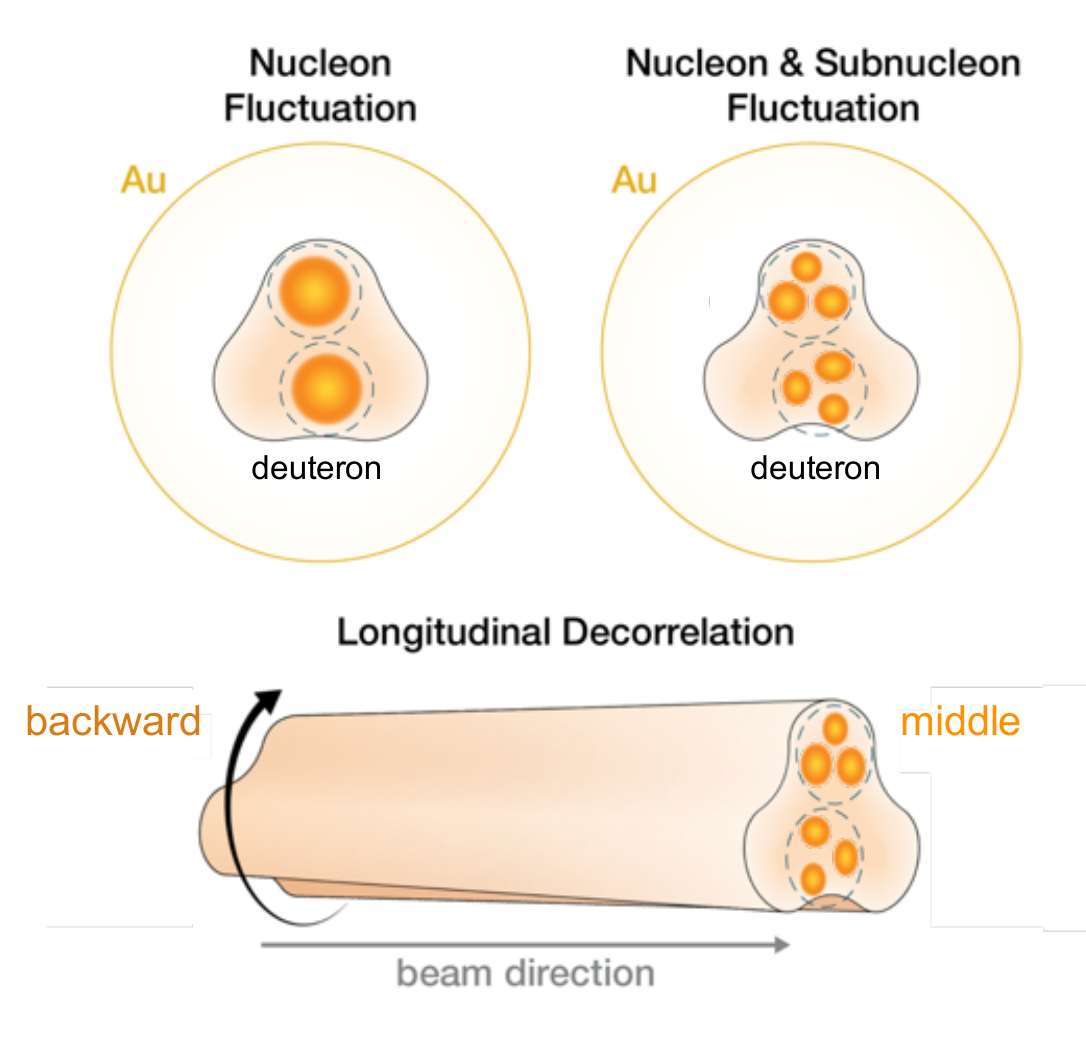}
\caption{Cartoon illustrating the interplay of three potential sources contributing to the triangular eccentricity $\varepsilon_3$ in asymmetric collisions like \dau: fluctuations in nucleon position (top-left), fluctuations in nucleon position along with their quark and gluon constituents (top-right), and fluctuations of the initial geometry defined by the overlap between deuteron and Gold nuclei along the beam direction, commonly referred to as longitudinal decorrelations (bottom).}\label{fig:1}
\end{figure}

One reason for encountering this challenge lies in the lack of quantitative control over the initial conditions and the associated $\varepsilon_n$. A crucial factor is whether each projectile nucleon should be considered a single smooth blob or as multiple blobs comprising gluon fields (as illustrated in the top panels of Fig.~\ref{fig:1}). Notably, flow data from \pp\ collisions at the LHC cannot be explained without invoking significant spatial fluctuations at the subnucleon level, which necessitates considering multiple distinct ``hot spots'' within each nucleon~\cite{Mantysaari:2016ykx}. These subnucleonic fluctuations are expected to be important in asymmetric collision systems like $p$+A or $d$+A collisions, although their $\snn$ dependencies remain unknown.

In \pau, \dau, and \heau\ collisions within the RHIC small system scan, the $\varepsilon_n$ values naturally depend on the assumed structure of the projectile $p$, $d$, and $^3$He, respectively. Table~\ref{tab:1} shows that the differences of $\varepsilon_3$ among these systems are particularly sensitive to whether each projectile nucleon is treated to follow a single smooth distribution or fluctuating distributions with a varying pattern from nucleon to nucleon. When nucleons are modeled as single smooth blobs, the $\varepsilon_3$ values in \pau\ and \dau\ collisions are significantly smaller than those in \heau\ collisions~\cite{Nagle:2013lja}. Conversely, considering each nucleon as three spatially separated blobs around valence quarks yields larger, yet much closer, $\varepsilon_3$ values for the three collision systems~\cite{Welsh:2016siu}. The impact of considering subnucleon-level fluctuations on $\varepsilon_3$ in \dau\ collisions is depicted in the top panels of Fig.~\ref{fig:1}. 

Another crucial aspect of the initial condition that introduces significant uncertainty is its longitudinal structure (as depicted in the bottom panel of Fig.~\ref{fig:1}). Experimental measurements in Pb+Pb, Xe+Xe, and $p$+Pb collisions at the LHC~\cite{CMS:2015xmx,ATLAS:2017rij,ATLAS:2020sgl}, along with supporting model studies~\cite{Bozek:2010vz,Jia:2014ysa,Bozek:2015bna,Pang:2015zrq,Shen:2017bsr,Bozek:2017qir,Zhao:2022ayk}, have revealed significant fluctuations in the shape of the initial geometry along the $\eta$ direction within each event. These fluctuations lead to significant decorrelation of the eccentricity vector as a function of $\eta$. Consequently, the extracted $v_n$ values from the two-particle correlation method depend on the chosen $\eta$ ranges of the particle pairs. A larger $\eta$ gap results in a smaller extracted $v_2$ signal. The decorrelation effect is more pronounced for $v_3$ and is particularly notable in smaller collision systems~\cite{CMS:2015xmx,Zhao:2022ugy}.

Distinguishing the effects of fluctuations at the nucleon and subnucleon levels, as well as those arising from longitudinal decorrelations in collision geometry, is imperative to establishing the creation of QGP in these small systems and extracting its properties.

To understand the origin of collectivity in small systems, particularly the role of collision geometry, RHIC has undertaken a scan of \pau, \dau, and \heau\ collisions. The PHENIX Collaboration measured $v_2$ and $v_3$ through correlations between particles in the central rapidity region and the backward (Au-going) rapidity region~\cite{PHENIX:2018lia, PHENIX:2021ubk}. The pseudorapidity gap $\Delta\eta$ ranges from three units to one unit, depending on the method used. The results reveal a hierarchy $v_{3}^{\scriptsize p\rm{Au}}\approx{v_{3}^{\scriptsize d\rm{Au}}}\approx \frac{1}{3}v_{3}^{\scriptsize  ^3\rm{HeAu}}$, consistent with model calculations employing a version of nucleon Glauber initial conditions~\cite{Nagle:2013lja}~\footnote{This calculation uses $\lr{\varepsilon_{n}}$ instead of the more correct $\sqrt{\lr{\varepsilon_{n}^2}}$, leading to a larger hierarchical differences, as shown in Table~\ref{tab:1}.}. Recently, STAR also measured $v_2$ and $v_3$ using correlations of particles closer to mid-rapidity while requiring a $\Delta\eta$ gap of one unit~\cite{STAR:2022pfn}. The findings suggest similar values of $v_3$ at comparable particle multiplicities in the three collision systems. The $v_{3}$ values in \heau\ are comparable between the two experiments, yet they differ notably in \pau\ and \dau. According to recent model calculations, these discrepancies are partly due to reduced longitudinal decorrelation in the STAR measurements~\cite{Zhao:2022ugy}. Resolving this issue experimentally requires direct determination of $v_3$ over a wide, continuous $\eta$ acceptance, which will be addressed in upcoming measurements utilizing recent STAR upgrades.

Another operational difference between the two experiments is that PHENIX did not perform an explicit non-flow subtraction. The rationale is that the non-flow component is reduced due to the large pseudorapidity gap between the middle and backward detectors, and any residual non-flow contributions are then covered by systematic uncertainties~\cite{PHENIX:2018lia}. Conversely, in the STAR analysis, larger non-flow contamination is expected owing to its smaller pseudorapidity gap, necessitating a careful estimation and subsequent subtraction of non-flow contributions~\cite{STAR:2022pfn}.

Our paper presents a detailed description of the procedures and non-flow subtraction methods used to extract $v_2$ and $v_3$ in these small systems. The effectiveness of these methods is validated through a thorough closure test of non-flow correlations within the HIJING model. A detailed study of the $\Deta$ dependence of $v_n$ is conducted to provide insights into the nature of flow decorrelations within the STAR acceptance. Additionally, an extensive comparison with hydrodynamic model calculations is performed, and prospects for future measurements are discussed.

\section{Data and Event Activity Selection}\label{sec:2}
\subsection{Event Selection}\label{sec:2.1}
The datasets used for this analysis include \pp, \pau, \dau, and \heau\ collisions at a center-of-mass energy of \TopE, collected by the STAR experiment during the years 2014, 2015, and 2016. Minimum Bias (MB) triggers are employed for data collection in both \pp\ and \heau\ collisions, while \pau\ and \dau\ collisions utilize both MB and High Multiplicity (HM) triggers.

The MB triggers in \pp, \pau, and \dau\ collisions require a coincidence between the east and west Vertex Position Detectors (VPD)~\cite{Llope:2014nva}, covering a rapidity range of 4.4 $< |\eta| <$ 4.9. For \heau\ collisions, the MB triggers require coincidences among the east and west VPD and the Beam-Beam Counters (BBC)~\cite{Bieser:2002ah}, along with at least one spectator neutron in the Zero Degree Calorimeter (ZDC)~\cite{Adler:2000bd} on the Au-going side. The rapidity coverage of these detectors is $3.3 < |\eta| < 5.1$ and $\eta < -6.5$, respectively. The MB trigger efficiency ranges from 60\% to 70\% for \pau, \dau, and \heau\ collisions systems. For MB \pp\ collisions, this efficiency was estimated to be around 36\%~\cite{STAR:2012nbd}.

In \pau\ and \dau\ collisions, the HM triggers require a minimum number of hits in the Time Of Flight (TOF) detector~\cite{Llope:2012zz}, in conjunction with the MB trigger criteria.

For offline analysis, events are selected based on their collision vertex position $z_{\mathrm{vtx}}$ relative to the Time Projection Chamber (TPC) center along the beam line. The chosen position falls within 2 cm of the beam spot in the transverse plane. The specific $z_{\mathrm{vtx}}$ ranges are optimized for each dataset, guided by distinct beam tuning conditions: 20 cm, 30 cm, 15 cm, and 30 cm for \pp, \pau, \dau, and \heau\ data, respectively. Moreover, to suppress pileup and beam background events in the TPC, a selection based on the correlation between the number of tracks in the TPC and those matched to the TOF detector is applied.

\subsection{Track Reconstruction and Selection}\label{sec:2.1b}
Charged particle tracks are reconstructed within $|\eta| < 1$ and $\pT > 0.2$ GeV/$c$ by the TPC. Track quality adheres to established STAR analysis standards: tracks are required to have at least 16 fit points in the TPC (out of a maximum of 45), with a fit-point-to-possible-hit ratio exceeding 0.52. To minimize contributions from secondary decays, tracks must have a distance of closest approach (DCA) to the primary collision vertex of less than 2 cm. Additionally, a valid track must be associated with a hit in the TOF detector or a signal in at least one strip layer in the Heavy Flavor Tracker (HFT) detector~\cite{Szelezniak:2015xC}. The TOF and HFT detectors provide faster response times than the TPC, effectively mitigating the effects of pileup tracks associated with multiple collisions accumulating during TPC drift time. To ensure high track reconstruction efficiency, only tracks within $|\eta| < 0.9$ are utilized in the correlation analysis.

\begin{figure*}[htbp]
\includegraphics[width=0.8\linewidth]{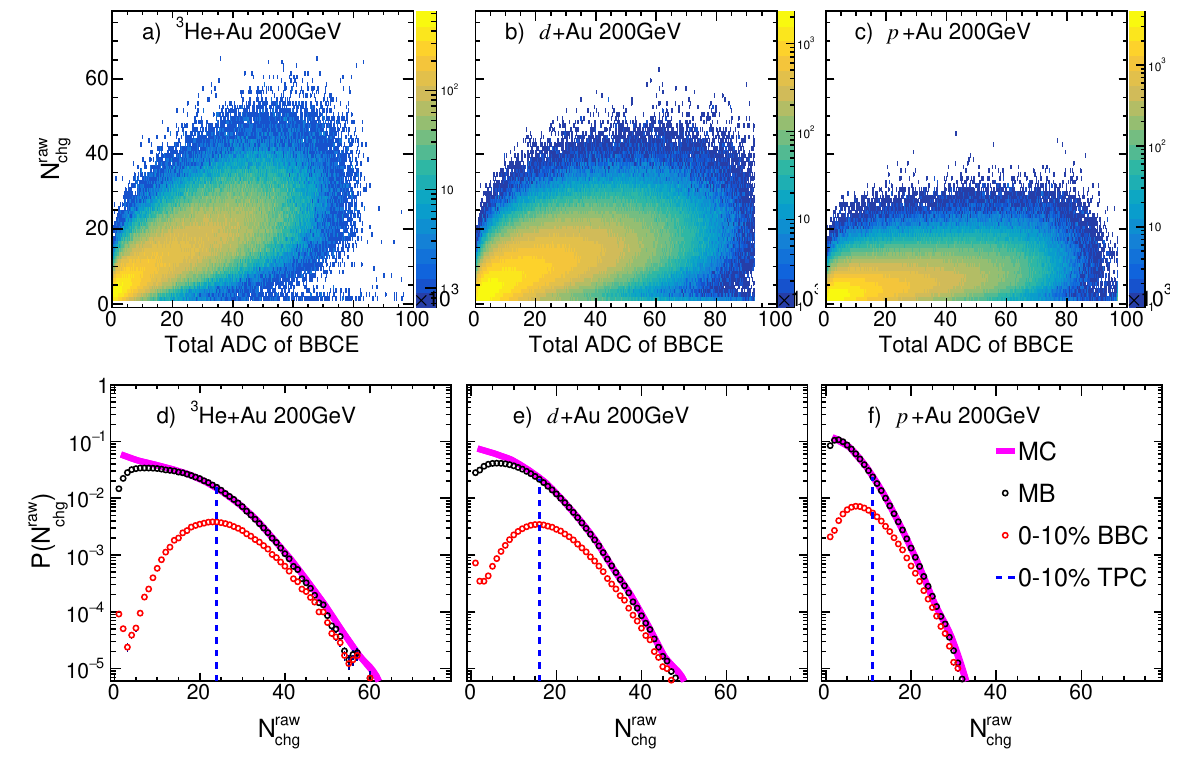}
\caption{Top row: Plot depicting $N_{\mathrm{ch}}^{\mathrm{raw}}$ vs. $\Sigma{Q_{\mathrm{BBCE}}}$ in minimum bias (\textsc{MB}) \heau, \dau, and \pau\ collisions at \TopE. The tiny fraction of events ($\ll1$\%), appearing as horizontal bands at very low $N_{\mathrm{ch}}^{\mathrm{raw}}$, originate fr2om the downstream beam related background. These events do not affect this analysis. Bottom row: Distribution of $N_{\mathrm{ch}}^{\mathrm{raw}}$ in each system. The black circles, red circles, and blue dashed lines correspond to \textsc{MB}, top 0--10\% event activity selected from BBCE and TPC, respectively. The pink solid curves indicate the generated multiplicity distribution derived from a Monte Carlo Glauber model fit (see text). }\label{fig:2}
\end{figure*}
The track reconstruction and matching efficiency are evaluated using the established STAR embedding technique~\cite{STAR:2009sxc}. This technique involves generating charged particles within a Monte Carlo generator, and subsequently subjecting them to a GEANT model representation of the STAR detector. The simulated detector signals are then merged with real data to capture the effects of the actual detector occupancy conditions. Subsequently, these merged events are reconstructed using the same offline reconstruction software for real data production.

The tracking efficiency is assessed by comparing the reconstructed tracks with the simulated input tracks. Specifically, tracking efficiency within the TPC exhibits minimal dependence on $\pT$ for values exceeding 0.5 GeV/$c$, reaching a plateau at approximately 0.9 across all collision systems. Applying a requirement for matching to the TOF detector reduces this plateau value to approximately 0.74.

\subsection{Event Activity Selection}\label{sec:2.2}
We aim to measure harmonic flow in $p/d/^{3}$He+Au collision events with large charged particle multiplicity or event activity. To achieve this, events are categorized into percentile ranges known as centrality classes, based on their apparent multiplicity as detected by a specific instrument. The most central events, situated within the top 0--10\% or the top 0--2\% of the multiplicity distribution, are chosen for subsequent analysis and comparison.

The default centrality classes are defined by employing the observed charged track multiplicity, $N_{\mathrm{ch}}^{\mathrm{raw}}$, within the pseudorapidity region $|\eta| <$ 0.9 and transverse momentum range of 0.2 $< \pT <$ 3.0 GeV/$c$ in the TPC~\cite{Anderson:2003ur}. These charged particle tracks are required to have a matched hit in the TOF detector. A Monte Carlo Glauber model, along with one of two distinct assumptions about particle production, is used to simulate the multiplicity distribution, which is then fitted to the $N_{\mathrm{ch}}^{\mathrm{raw}}$ to determine the centrality percentiles.

The first approach is based on the two-component model for particle production~\cite{Kharzeev:2000ph}, where the number of sources for particle production is assumed to be
\begin{equation}
N_s =\left[ (1-x) \frac{N_{\rm part}}{2} + xN_{\rm coll} \right],
\end{equation}
where $x$ is the fraction of the second component. The number of participants, $\npart$, and number of collisions, $N_{\rm coll}$, are extracted from the PHOBOS Glauber Monte Carlo simulation~\cite{PHOBOSGlauber}. In the second approach, the $N_{s}$ is assumed to follow a power law dependence on $\npart$, $N_{s}=\npart^{\alpha}$. 

Multiplicity fluctuation is incorporated via the Negative Binomial Distributions (NBD) for each source,
\begin{equation}
P_{\rm NBD}(\mu, k; n) = \frac{\Gamma(n+k)}{\Gamma(n+1)\Gamma(k)} \cdot \frac{(\mu/k)^n}{(\mu/k + 1)^{n+k}},
\end{equation}
where $n$ is the generated multiplicity, and $\mu$ and $k$ are free parameters. The inefficiency for triggering events with a single source is denoted by $\varepsilon$. 

The multiplicity of an event at the generator level $N_{\mathrm{ch}}^{\mathrm{mc}}$ is obtained by summing $n$ for all $N_s$ sources.  The corresponding multiplicity after accounting for trigger inefficiency, denoted by $N_{\mathrm{ch}}^{\mathrm{obs}}$, is also obtained.

The distribution of $N_{\mathrm{ch}}^{\mathrm{obs}}$ is then fitted to measured $N_{\mathrm{ch}}^{\mathrm{raw}}$ distributions for each collision system. The trigger inefficiency $\varepsilon$ and NBD parameters $\mu$ and $k$ are adjusted to achieve an optimal global fit. This procedure also yields a multiplicity distribution at the generator level, $N_{\mathrm{ch}}^{\mathrm{mc}}$, from which we can determine the centrality percentiles. 

Examples of $N_{\mathrm{ch}}^{\mathrm{mc}}$ from the first approach are displayed in the lower panels of Fig.~\ref{fig:2} for the three collision systems. The apparent deviations at low $N_{\mathrm{ch}}^{\mathrm{raw}}$ values are attributable to the inefficiency of the MB triggers, while the simulated distribution agrees with the data at large $N_{\mathrm{ch}}^{\mathrm{raw}}$ values. The values of $\lr{N_{\mathrm{ch}}^{\mathrm{raw}}}$ are found to be slightly different between the two approaches. For the top 0--10\% centrality interval, they amount to a 4\% difference in \pau\ collisions and 3\% in $d$/$^3$He+Au collisions.

To examine the potential auto-correlation between event selection and flow signal, an alternative event activity selection is introduced as a cross-check. This selection relies on the signal from the BBC on the Au-going side (denoted as BBCE) within a pseudorapidity range of $-5.0 < \eta < -3.3$. For instance, the 0--10\% event classes are characterized as the top 10\% of the total charge registered by the BBCE, denoted as $\Sigma Q_{\rm BBCE}$. The correlation between $N_{\mathrm{ch}}^{\mathrm{raw}}$ and $\Sigma Q_{\rm BBCE}$ is illustrated in the upper panels of Fig.~\ref{fig:2} for MB \pau, \dau, and \heau\ collisions. A broad correlation is observed in all three systems, implying that events in a narrow range of $N_{\mathrm{ch}}^{\mathrm{raw}}$ can have a large spread in $\Sigma Q_{\rm BBCE}$ and vice versa. Corresponding $N_{\mathrm{ch}}^{\mathrm{raw}}$ distributions for MB and 0--10\% events, selected via TPC and BBC, are displayed in the lower panels.

Table~\ref{tab:2} provides the efficiency-corrected multiplicities, $\lr{N_{\mathrm{ch}}}$, for MB \pp\ and the 0--10\% most central $p$/$d$/\heau\ collisions, selected using both $N_{\mathrm{ch}}^{\mathrm{raw}}$ and $\Sigma Q_{\rm BBCE}$. Additionally, the table presents values for the 0--2\% most central $p$/$d$+Au collisions, selected with TPC-based centrality. The systematic uncertainties on $\lr{N_{\mathrm{ch}}}$ arise mainly from uncertainties in charged pion reconstruction efficiency, evaluated through the embedding procedure. The additional PID dependence of the reconstruction efficiency associated with $K^{\pm}$ and (anti-)protons are estimated from embedding and the known particle ratios~\cite{STAR:2008med}. The total uncertainty associated with the efficiency correction is estimated to be around 5\%.

Note that the $\lr{N_{\mathrm{ch}}}$ values quoted for MB \pp\ collisions are not corrected for the trigger inefficiency and, therefore, should be treated as the value for selected events.

\begin{table}[htbp]
   \begin{center}
   \begin{tabular}{cc|ccc}
\hline \hline
& MB \pp & \pau & \dau & \heau \\
     \hline
\multirow{6}{*}{$\left<N_{\mathrm{ch}}\right>$} & \multirow{6}{*}{4.7$\pm$0.3} 
    & \multicolumn{3}{ c }{0--10\% from TPC}\\
    &  & \multicolumn{1}{ c }{21.9$\pm$1.1} & 35.6$\pm$1.8 & 47.7$\pm$2.4 \\\cline{3-5}
     & & \multicolumn{3}{ c }{0--2\% from TPC}\\
     & & \multicolumn{1}{ c  }{ 34.1$\pm$1.7} & 46.4$\pm$2.3 & - \\
     \cline{3-5}
     & &\multicolumn{3}{ c }{0--10\% from BBC}\\
     & & \multicolumn{1}{ c  }{15.7$\pm$0.8} & 27.6$\pm$1.4 & 41.6$\pm$2.1 \\\cline{3-5}
     \hline \hline
    \end{tabular}
\end{center}
\caption{\label{tab:2} The efficiency-corrected average multiplicity, $\lr{N_{\mathrm{ch}}}$, for MB \pp,  0--10\% most central $p$/$d$/\heau\ collisions, as well as 0--2\% most central \pau\ and \dau\ collisions using TPC-based centrality. The values obtained for 0--10\% BBC-based centrality are also shown.}
\end{table}

\section{Methodology for $v_{n}$ extraction}\label{sec:3}
\subsection{Two-particle correlation function and per-trigger yield}\label{sec:3.1}
\begin{figure*}[htbp]
\includegraphics[width=0.75\linewidth]{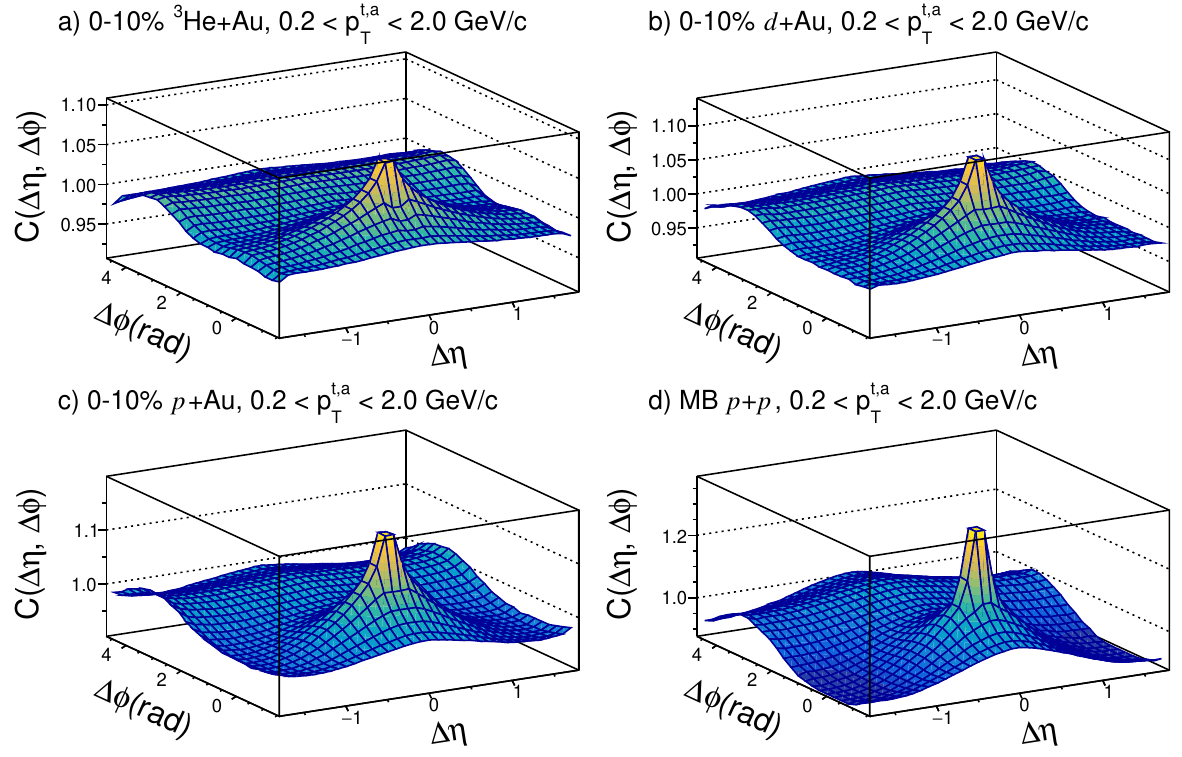}
\caption{The normalized two-particle correlation function presented as a function of \Deta\ and \Dphi\ for the trigger and associated particles within the 0.2 $<\pT<$ 2.0 GeV/$c$ range in central \pau, \dau, \heau, and \textsc{MB} \pp\ collisions at \TopE.}\label{fig:3}
\end{figure*}
 \begin{figure*}[htbp]
\includegraphics[width=1\linewidth]{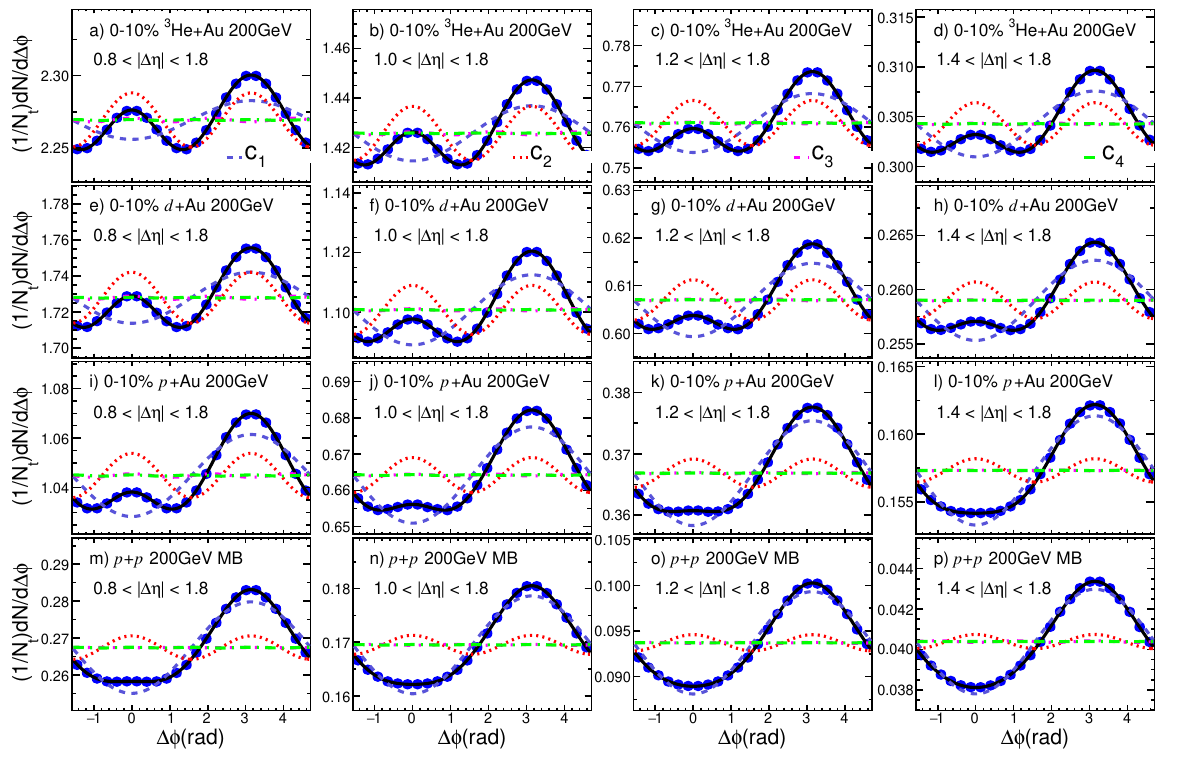}
\caption{The per-trigger yield displayed as a function of \Dphi\ in central \heau\ (top row), \dau\ (second row), \pau\ (third row), and \textsc{MB} \pp\ (bottom row) collisions for particles with $0.2 < \pT^{\mathrm t},\pT^{\mathrm a} < 2.0$~GeV/$c$. The plots are provided for four different \Deta\ selections, moving from left to right. The color curves represent the Fourier components obtained from the Fourier expansion of the per-trigger yield.}\label{fig:4}
\end{figure*}

This analysis measures two-particle correlations as functions of the relative pseudorapidity, $\Delta\eta$, and relative azimuthal angle, $\Delta\phi$~\cite{PHENIX:2008osq}. Trigger particles are defined as charged particle tracks within $|\eta| < 0.9$ and $0.2 < \pT^{\mathrm t} < 2.0$ GeV/$c$. Pairs of particles are then formed by pairing each trigger particle with the remaining charged particle tracks that satisfy $|\eta| < 0.9$ and $0.2 < \pT^{\mathrm a} < 2.0$ GeV/$c$. This leads to a maximum gap of $|\Delta\eta| < 1.8$ between the pairs. The track reconstruction efficiency is applied to individual particles.

The two-dimensional two-particle correlation function, $C(\Delta\eta,\Delta\phi)$, is calculated as:
\begin{align}
 C(\Delta\eta,\Delta\phi) = \frac{\int{B(\Delta\eta',\Delta\phi')d\Delta\phi' d\Delta\eta'}}{\int{S(\Delta\eta', \Delta\phi')d\Delta\phi' d\Delta\eta'}}\frac{S(\Delta\eta,\Delta\phi)}{B(\Delta\eta,\Delta\phi)}\;,
 \label{eq:4}
\end{align}
where $S(\Delta\eta,\Delta\phi)$ and $B(\Delta\eta,\Delta\phi)$ represent the pair distributions from same-event and mixed-event samples, respectively. Mixed-event pairs are formed by combining tracks from two different events with similar centrality and similar $z_{\mathrm{vtx}}$, as detailed in Ref.~\cite{PHENIX:2008osq}.

Correlation functions $C(\Delta\eta,\Delta\phi)$ are obtained for different collision systems with centrality selection based on the TPC multiplicity. The resulting correlation functions from MB events are displayed in Fig.~\ref{fig:3} for $0.2 < \pT^{\mathrm t} < 2.0$ GeV/$c$ (examples for other $\pT^{\mathrm t}$ ranges are shown in Appendix~\ref{sec:app}). A ridge-like structure around $\Delta\phi=0$ and along the $\Delta\eta$ direction is clearly observed in central \dau\ and $^3$He+Au collisions, and possibly in \pau\ collisions, whereas it is absent in MB \pp\ collisions.

One-dimensional correlation functions, $C(\Delta\phi)$, are obtained as:
\begin{eqnarray}
 C(\Delta\phi) = \frac{\int{B(\Delta\phi')d\Delta\phi'}}{\int{S(\Delta\phi')d\Delta\phi'}}\frac{S(\Delta\phi)}{B(\Delta\phi)}\;,
 \label{eq:5}
\end{eqnarray}
where $S(\Delta\phi)$ and $B(\Delta\phi)$ are obtained by integrating $S(\Delta\eta,\Delta\phi)$ and $B(\Delta\eta,\Delta\phi)$ using four distinct ranges of $|\Delta\eta|$: $|\Delta\eta|>$ 0.8, 1.0, 1.2, and 1.4. The per-trigger yield, denoted as \Yphi, is defined as:
\begin{eqnarray}
 Y(\Delta\phi) \equiv \frac{1}{N_{\mathrm t}}\frac{dN}{d\Delta\phi} =\frac{C(\Delta\phi)\int{S(\Delta\phi)d\Delta\phi}}{N_{\mathrm t}\int{d\Delta\phi}}\;,
\label{eq:6}
\end{eqnarray}
where $N_{\mathrm t}$ is the number of trigger particles after efficiency correction.

Figure~\ref{fig:4} illustrates $Y(\Delta\phi)$ obtained for MB \pp\ and the 0--10\% most central \pau, \dau, and $^3$He+Au collisions in the four $|\Delta\eta|$ ranges, with $\pT$ of trigger particles in $0.2 < \pT^{\mathrm t} < 2.0$ GeV/$c$. The correlation functions for other $\pT^{\mathrm t}$ ranges can be found in Figs.~\ref{fig:app2}--\ref{fig:app5} in Appendix~\ref{sec:app}.

After a gap cut of $\Deta_{\mathrm{min}}<|\Delta\eta|<1.8$ to suppress non-flow, with $\Deta_{\mathrm{min}} = 0.8$, 1.0, 1.2, or 1.4 as shown in Fig.~\ref{fig:4}, prominent near-side peaks are observed in central \dau\ and $^3$He+Au collisions. These near-side peaks may be attributed to contributions from long-range collective flow. Meanwhile, the large away-side peaks are predominately attributed to the non-flow correlations from dijet fragmentations. In contrast, MB \pp\ correlation functions exhibit weak near-side peaks but much stronger away-side peaks, suggesting that non-flow contributions dominate the entire correlation structure. Hence, the \pp\ data provide a baseline for assessing the remaining non-flow contributions in $p$/$d$/$^3$He+Au collisions.

The main goal of the gap cut is to suppress the significant near-side jet peaks observed in Fig.~\ref{fig:3}. In \pp\ collisions, however, the near-side of the correlation function still exhibits a low-amplitude, broad peak for $\Deta_{\mathrm{min}}=0.8$, which decreases for larger gap cuts. For this analysis, a default $\Deta_{\mathrm{min}}=1.0$ gap cut is chosen in all four systems, which achieves a reasonable suppression of the near-side jet peak while still maintaining decent statistical precision. More details can be found in Sec.\ref{sec:3.3}.

\subsection{Non-flow subtraction and $v_n$ extraction}\label{sec:3.2}
This section presents four non-flow subtraction methods,, outlining their fundamentals, similarities, differences, and performance across various collision systems.

All methods begin with the Fourier decomposition of the one-dimensional per-trigger yield distribution, \Yphi\:
\begin{eqnarray}
 Y(\Delta\phi) = c_{0}(1 + \sum_{n=1}^{4} \, 2c_{n}\, \cos ( n \,\Delta\phi )) \;,
 \label{eq:nf0}
\end{eqnarray}
where $c_{0}$ represents the average pair yield (the pedestal), and $c_n$ (for $n=1$ to 4) are the Fourier coefficients. The corresponding harmonic components are depicted by the colored dashed lines in Fig.~\ref{fig:4}.

The $c_n$ values are influenced by non-flow correlations, particularly on the away side, which need to be estimated and subtracted. The four established methods for estimating non-flow are:
\begin{enumerate}
\item the $c_0$ method~\cite{ALICE:2012eyl,Aad:2012gla,PHENIX:2013ktj}.
\item the near-side subtraction method~\cite{STAR:2004jwm,ATLAS:2014qaj,CMS:2016fnw,Lim:2019cys}.
\item the $c_1$ method~\cite{PHENIX:2017nvb}.
\item the template-fit method~\cite{ATLAS:2015hzw,ATLAS:2018ngv}. 
\end{enumerate}
A discussion of some of the methods can be found in Refs.~\cite{Lim:2019cys}.

\begin{figure*}[htbp]
\includegraphics[width=1\linewidth]{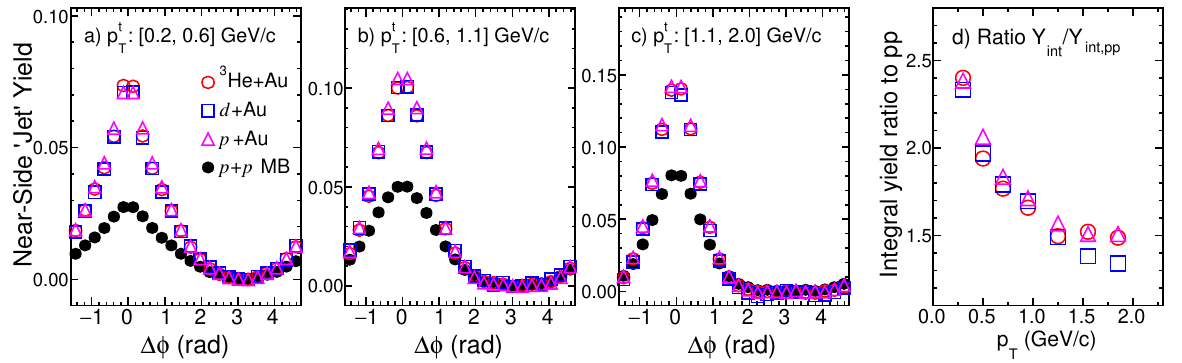}
\caption{Estimated yield of near-side jet-like correlations for different trigger $\pT$ values in \textsc{MB} \pp\ and the top 0--10\% of \pau, \dau, and \heau\ collisions at \TopE, distributed across three $\pT^{\mathrm t}$ ranges in panels a), b), and c), respectively. In panel d), the ratios of yield between $p$/$d$/$^{3}$He+Au and \textsc{MB} \pp, $Y_{\mathrm{int}}/Y_{\mathrm{int},pp}$, are demonstrated as a function of trigger $\pT$. Associated particles are selected from $0.2<\pT^{\mathrm a}<2$ GeV/$c$.}
\label{fig:5}
\end{figure*}
\begin{figure*}[htbp]
\includegraphics[width=1\linewidth]{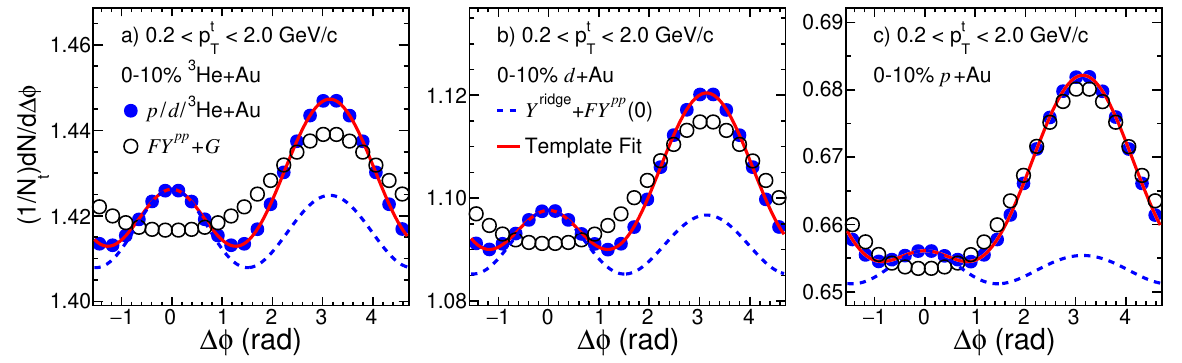}
\caption{Performance assessment of the template-fit method for triggers within 0.2$<\pT<$2.0 GeV/$c$ in the top 0--10\% of \pau, \dau, and \heau\ collisions at \TopE. This assessment includes the ridge yield ($\YphiRidge$) and the scaled yield of \pp\ collisions ($\Yphipp$) multiplied by a factor of $F$. The coefficient $G$ represents the integrated yield of $\YphiRidge$. Further details can be found in Eq.~\ref{eq:nf5}, Eq.~\ref{eq:nf6}, and the accompanying text.}
\label{fig:6}
\end{figure*}

{\bf $c_0$ Method}. In the $c_0$ method, non-flow effects in $p$/$d$/$^3$He+Au collisions are assumed to arise from a convolution of several independent \pp\ collisions, expected to be proportional to $c_n^{pp}/c_0$. The Fourier coefficients, after subtracting non-flow contributions, are calculated as:
\begin{align}
c_{n}^{\mathrm {sub}}  =  c_{n} - \frac{c_{0}^{pp}}{c_{0}}\times c_{n}^{pp}\;.
\label{eq:nf1}
\end{align}
%This method is similar to the ``scalar-product method'' mentioned in Refs.~\cite{STAR:2004amg} and \cite{STAR:2004jwm}.

{\bf Near-side Subtraction Method}. The $c_0$ method may underestimate non-flow contributions in central $p$/$d$/$^3$He+Au collisions due to the selection bias of high-multiplicity events. The near-side subtraction method addresses this bias by estimating differences in non-flow contributions between \pp\ and $p$/$d$/$^3$He+Au collisions using the near-side per-trigger yield, $Y^{\mathrm N}(\Delta\phi)$. It is defined as the difference between the short-range yield integrated over $0.2 < |\Delta\eta| < 0.5$, $Y^{\mathrm S}(\Delta\phi)$, and the long-range yield integrated over $1.0 < |\Delta\eta| < 1.8$, $Y^{\mathrm L}(\Delta\phi)$:
\small{\begin{align}\label{eq:nf2}
Y_{\mathrm {int}} \equiv \int\!{Y^{\mathrm N}d\Dphi}=\!\int\!{(Y^{\mathrm S}-fY^{\mathrm L})d\Dphi}\;,
\end{align}}\normalsize
where $f=\frac{Y^{\mathrm S}(\Dphi=\pi)}{Y^{\mathrm L}(\Dphi=\pi)}$. The non-flow subtracted Fourier coefficients are then:
\begin{align}
c_{n}^{\mathrm {sub}} = c_{n} - \frac{Y_{\mathrm{int}}}{Y_{\mathrm{int},pp}}\frac{c_{0}^{pp}}{c_{0}}\times c_{n}^{pp},
\label{eq:nf3}
\end{align}

The $Y^{\mathrm N}(\Delta\phi)$ distributions for various trigger particle $\pT$ ranges are depicted in Fig.~\ref{fig:5}, and the ratio $Y_{\mathrm{int}}/Y_{\mathrm{int},pp}$ is shown in the right panel of the same figure. This ratio starts around 2.4 at low $\pT$ and decreases rapidly with $\pT$ while staying above unity. This indicates that the near-side subtraction method, compared to the $c_0$ method, removes a much larger portion of \pp-scaled correlations attributed to non-flow.

{\bf $c_1$ Method}.  In this method, non-flow contributions are directly estimated from the away-side jet-like correlations. Here, the away-side jet contribution is assumed to scale with the $c_1$ component from the Fourier decomposition of $Y(\Delta\phi)$. This assumption holds at low $\pT$, where the away-side jet shape is well described by a $1+2c_1\cos(\Delta\phi)$ function. However, we find that it is also a valid assumption over the entire $\pT$ range considered in this analysis. Thus, the ratio of the non-flow component between \pp\ and $p$/$d$/$^3$He+Au is expected to be captured by the ratio of their respective $c_1$ values~\cite{PHENIX:2017nvb}. The non-flow subtracted Fourier coefficients are then calculated as,
\begin{align}
c_{n}^{\mathrm {sub}} =  c_{n} - \frac{c_{1}}{c_{1}^{pp}}\times c_{n}^{pp}\;.
\label{eq:nf4}
\end{align}

\begin{figure*}[t!]
\includegraphics[width=1.0\linewidth]{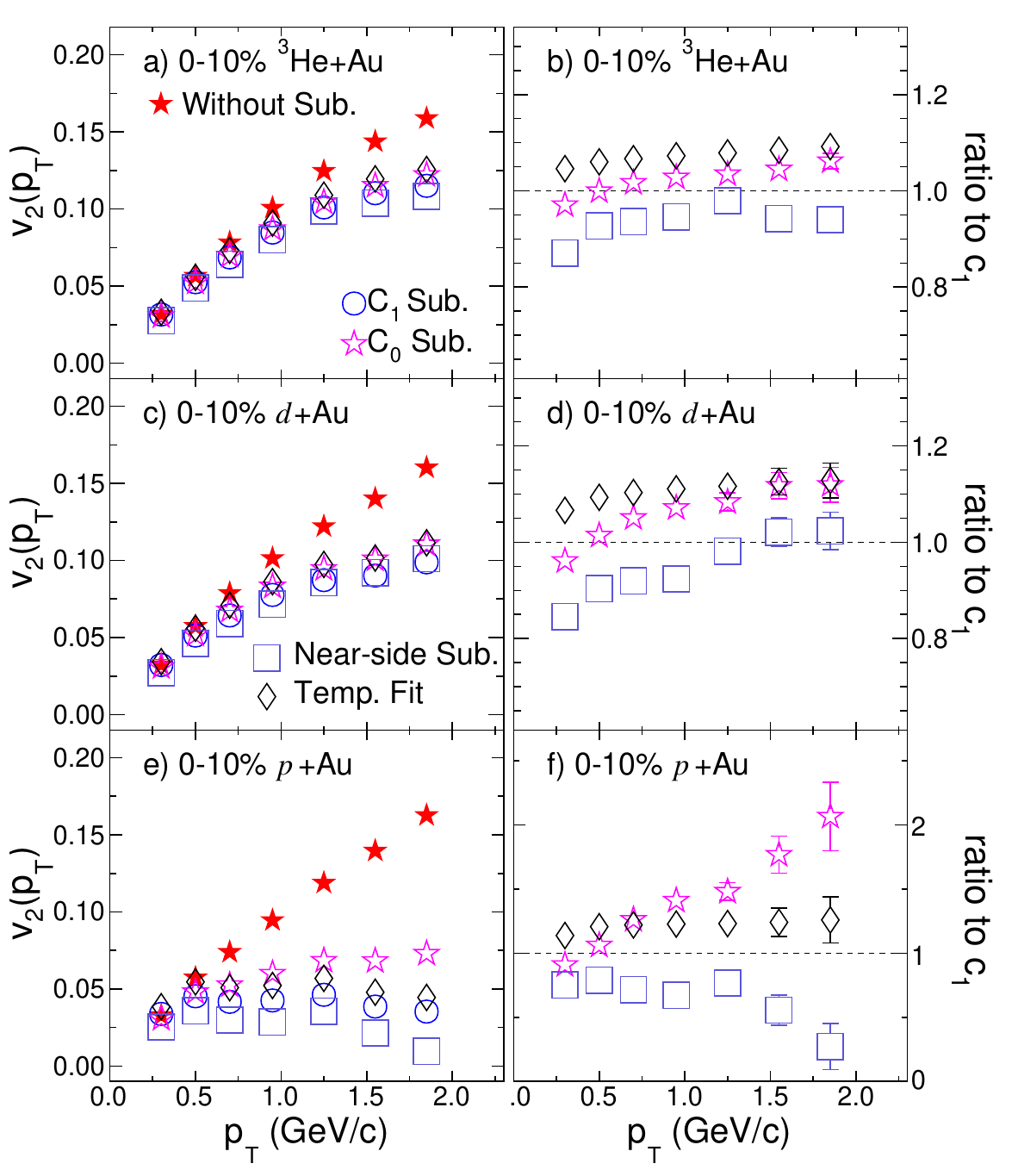}
\caption{The $v_2(\pT)$ values (left figure) and $v_3(\pT)$ values (right figure) derived from different non-flow subtraction methods (left column) along with the ratios compared to those obtained through the $c_1$ method (right column) in the 0--10\% most central \pau, \dau, and \heau\ collisions. Only statistical uncertainties are depicted.}
\label{fig:7}
\end{figure*}

\begin{figure*}[htbp]
\includegraphics[width=1\linewidth]{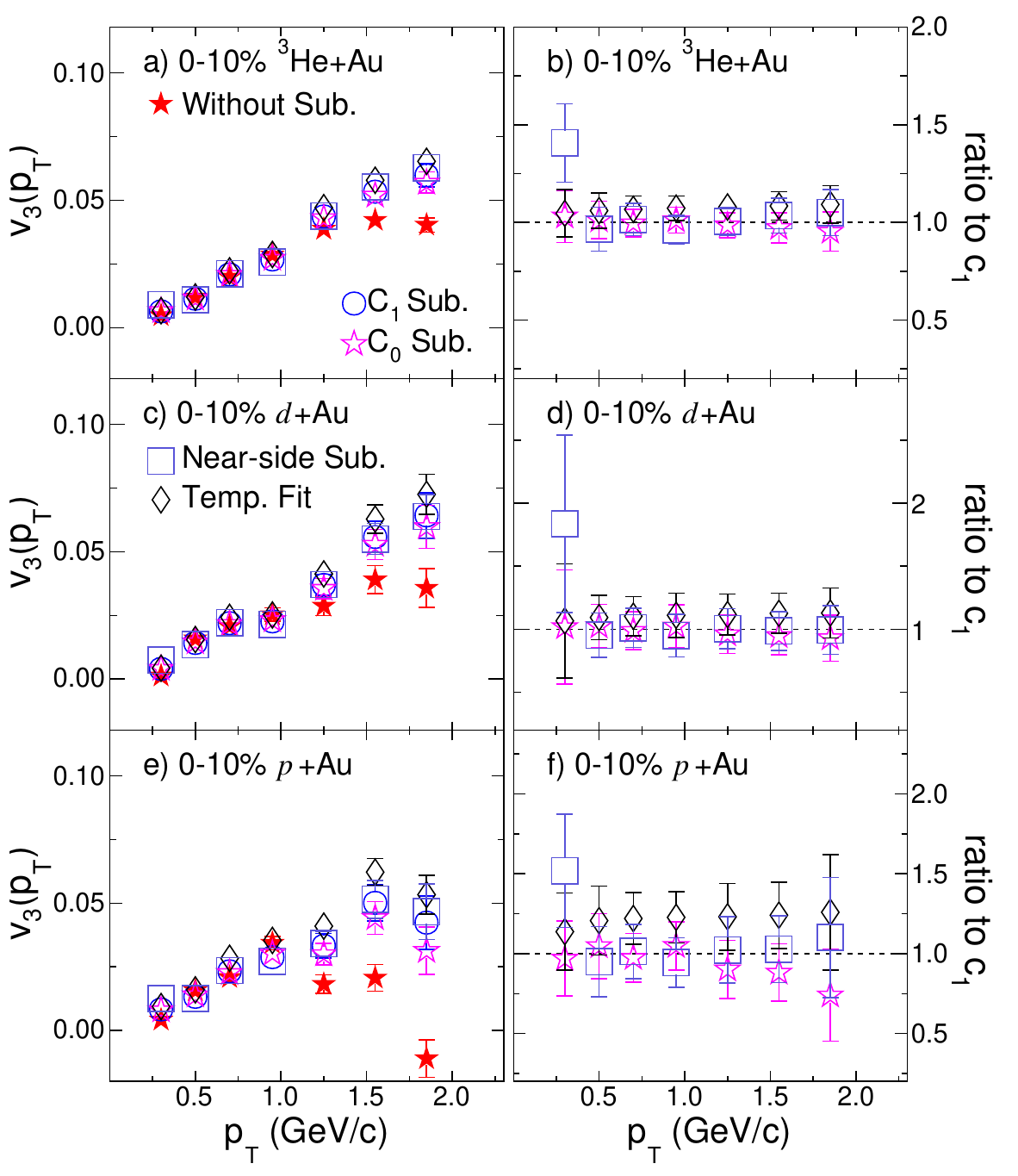}
\caption{The $v_2(\pT)$ values (left figure) and $v_3(\pT)$ values (right figure) derived from different non-flow subtraction methods (left column) along with the ratios compared to those obtained through the $c_1$ method (right column) in the 0--2\% most central \pau\ and \dau\ collisions. Only statistical uncertainties are depicted.}
\label{fig:8}
\end{figure*}

{\bf Template-fit Method}. This last method is developed by the ATLAS Collaboration and detailed in Ref.~\cite{ATLAS:2015hzw}. This method assumes that the \Yphi\ in $p$/$d$/$^3$He+Au collisions is a linear combination of a scaled \Yphi\ distribution from MB \pp\ collisions, representing all non-flow contributions, and a \Yphi\ distribution containing only genuine collective flow, denoted as $\YphiRidge$ ,
\begin{align}\label{eq:nf5}
\YphiTempl = F\Yphipp +  \YphiRidge\, ,
\end{align}
where 
\begin{align}
\YphiRidge = G \left(1 + 2\sum_{n=2}^{4}\ c_{n}^{\mathrm {sub}} \cos{(n\Delta \phi)}\right)\, .
\label{eq:nf6}
\end{align}
The parameters $F$ and $c_{n}^{\mathrm {sub}}$ are determined through fitting the data to $\YphiTempl$. The coefficient $G$, determining the magnitude of the pedestal of $\YphiRidge$, is fixed by ensuring that the integral of $\YphiTempl$ equals the integral of \Yphi\ . 

The performance of the template-fit method is shown in Fig.~\ref{fig:6}. The narrowing of the away-side peak in $p$/$d$/$^3$He+Au collisions compared to \pp\ collisions indicates the presence of a significant $\cos(2\Delta\phi)$ component~\cite{ATLAS:2015hzw}. Since both the $c_1$ method and the template-fit method use the away-side jet correlation to constrain non-flow, the scale factors in Eqs.~\ref{eq:nf4} and \ref{eq:nf5} are expected to be similar, i.e., $F\approx c_{1}/c_{1}^{pp}$. The primary difference lies in how they handle flow modulation: The $c_{1}$ method assumes that flow modulation affects all particle pairs, as captured by the $c_0$ term in Eq.~\ref{eq:nf0}, whereas the template-fit method assumes that flow modulation applies only to the subtracted pedestal, as represented by the parameter $G$ in Eq.~\ref{eq:nf6}. This implies that in central $p$/$d$/\heau\ collisions, where the particle multiplicity is much larger than that in \pp\ collisions, the template-fit method is almost identical to the $c_{1}$ subtraction method.

{\bf Comparison of Methods}. The scale factors obtained from the four non-flow subtraction methods, as given by Eqs.~\ref{eq:nf1}, \ref{eq:nf3}, \ref{eq:nf4}, and \ref{eq:nf5}, follow a consistent ordering: $\frac{c_{0}^{pp}}{c_{0}}< F\approx \frac{c_{1}}{c_{1}^{pp}}< \frac{Y_{\mathrm{int}}}{Y_{\mathrm{int},pp}}\frac{c_{0}^{pp}}{c_{0}}$. This indicates that the results obtained from the $c_1$ method and the template-fit method lie between those obtained from the $c_0$ method and the near-side subtraction method.

The difference in scale factors arises from the biases associated with jet fragmentation on the near side and the away side, varying across the subtraction methods. In two-particle correlations, pairs within the near-side jet peak require two particles originating from the same jet, while pairs within the away-side jet peak only need one particle each from the near-side and away-side jets. As a result, the near-side subtraction method tends to overestimate the non-flow contribution due to a larger jet fragmentation bias, while the $c_{0}$ method tends to underestimate the non-flow contribution. Based on this analysis, the $c_{1}$ method is chosen as the default method in this study.

Note that the MB \pp\ events used for non-flow estimation are biased towards higher multiplicity due to trigger efficiency. However, assuming that the shape of non-flow contribution in the correlation function remains unmodified, trigger inefficiency in \pp\ collisions is not expected to influence the subtraction procedure.

Finally, the flow coefficients $v_n$ are calculated using the two-particle harmonics $c_n(\pT^{\mathrm t},\pT^{\mathrm a})$ with or without non-flow subtractions:
\begin{align}\label{eq:nf7}
v_{n}(\pT^{\mathrm t}) = \frac{c_{n}(\pT^{\mathrm t},\pT^{\mathrm a})}{\sqrt{c_{n}(\pT^{\mathrm a},\pT^{\mathrm a})}}\;,
\end{align}
By default, particle pairs are required to have a pseudorapidity gap of $|\Deta|>1$, with associated particles in the range $0.2<\pT^{\mathrm a}<2$ GeV/$c$.

{\bf Results and Discussion} The left part of Fig.~\ref{fig:7} illustrates the extracted $v_2(\pT)$ in 0--10\% central \pau, \dau, and \heau\ collisions using different non-flow subtraction methods. Results agree with those before non-flow subtraction at low $\pT$ ($<0.6$ GeV/$c$), but are systematically smaller at higher $\pT$. This behavior aligns with the expectation that non-flow correlations from the away-side jet contribute more in smaller collision systems and at higher $\pT$. 

Among the four non-flow subtraction methods, $v_2(\pT)$ values agree within 20\% in \dau\ and \heau\ collisions. However, in \pau\ collisions, $v_2(\pT)$ values are similar at $\pT<0.6$ GeV/$c$ but exhibit a noticeable spread at higher $\pT$. This observation suggests that $v_2(\pT)$ values can be extracted up to 2 GeV/$c$ in \dau\ and \heau, but only up to 0.6 GeV/$c$ in \pau\ collisions.

The right part of Fig.~\ref{fig:7} presents the same comparison for $v_3(\pT)$. Results after non-flow subtraction closely resemble those obtained without non-flow subtraction up to 1 GeV/$c$, but are slightly larger at higher $\pT$. The overall impact of non-flow correlations on $v_3(\pT)$ is significantly smaller than on $v_2(\pT)$, resulting in a weaker dependence of $v_3(\pT)$ values on the non-flow subtraction methods. This is because the away-side jet correlation around $\Delta\phi\sim \pi$, being very broad within the considered $\pT^{\mathrm a}, \pT^{\mathrm t}$ range, gives rise to large negative $c_1$, a smaller positive $c_2$, and a much smaller negative $c_3$. The negative non-flow contribution to $c_3$ implies that non-flow subtraction can only increase $v_3$, as observed in Fig.~\ref{fig:7}. The spreads of $v_3(\pT)$ from different non-flow subtraction methods are approximately 10\% in \dau\ and \heau, increasing to 20--30\% in \pau\ collisions.

The same analysis is conducted for the 0--2\% ultracentral $p$/$d$+Au collisions, with results shown in Fig.~\ref{fig:8}. The dependence on the non-flow subtraction methods is qualitatively similar for both $v_2$ and $v_3$, although quantitatively, variations in \pau\ collisions are significantly reduced compared to Fig.~\ref{fig:7}. This reduction can be attributed to higher $\langle N_{\mathrm{ch}} \rangle$ values in the 0--2\% centrality range compared to the 0--10\% centrality range in \pau\ collisions. Larger $\langle N_{\mathrm{ch}} \rangle$ implies a significant decrease in scale factors in all the non-flow subtraction methods, such as $c_{0}^{pp}/c_{0}$ in the $c_0$ and near-side subtraction methods, $c_{1}/c_{1}^{pp}$ in the $c_1$ method, and the $F$ in the template-fit method. This reduction in the scale factors diminishes the sensitivity to non-flow correlations and leads to smaller variations among methods. This effect is most significant in \pau\ collisions and less pronounced in \dau\ collisions.

\begin{figure*}[htbp]
\centering
\includegraphics[width=0.85\linewidth]{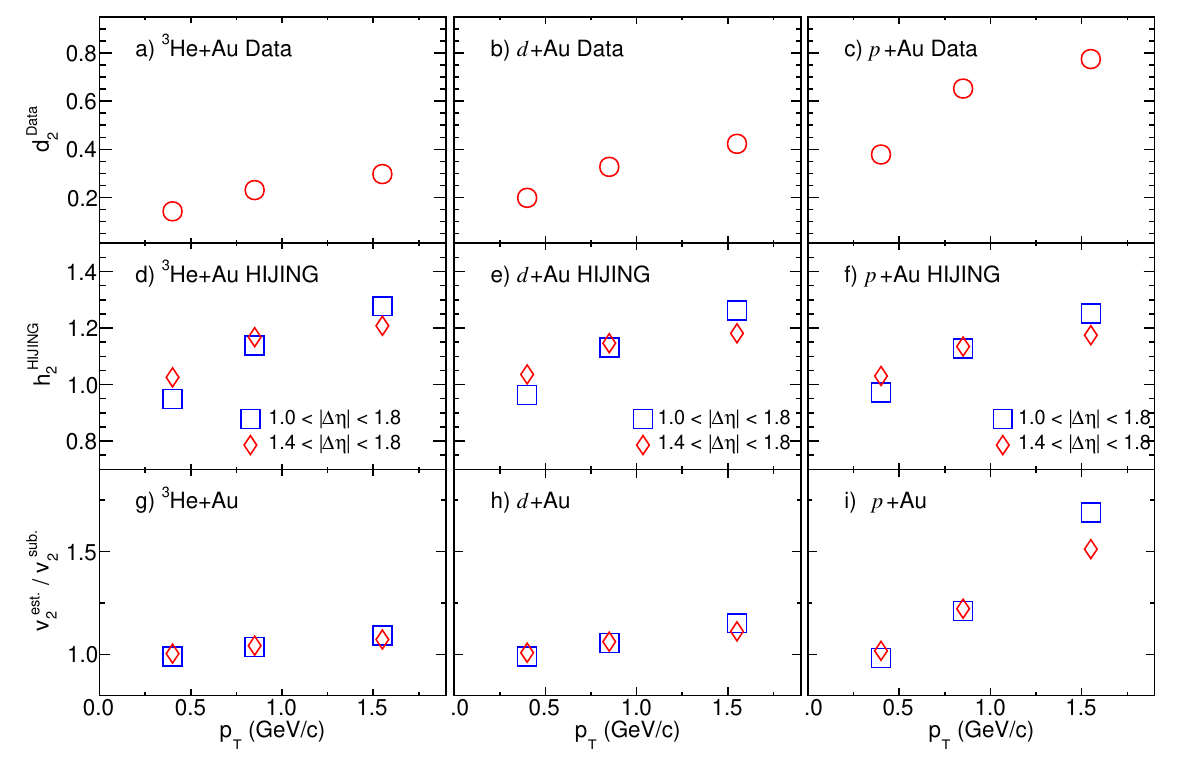}
%\caption{The values of $d_2 \equiv \frac{c_2^{pp}}{c_2} \frac{c_1}{c_1^{pp}}$ calculated from real data with $|\Delta \eta|>1$ (top row), $h_2 \equiv \frac{c_2^{pp \mathrm{hij}}}{c_2^{\mathrm{hij}}} \frac{c_1}{c_1^{pp}}$ derived from the HIJING model for two ranges of $|\Delta \eta|$ (middle row), and $v_2^{\text {est}} / v_2^{\text {sub}}$ defined in Eq.~\ref{eq:cl5} displayed as a function of $\pT$ in central \heau\  (left column), \dau\ (middle column), and \pau\ (right column) collisions. Only statistical uncertainties are shown.}\label{fig:12}
\caption{The values of $d_2$ defined in Eq.~\ref{eq:cl4} and calculated from real data with $|\Delta \eta|>1$ (top row), $h_2$ defined in Eq.~\ref{eq:cl2} and derived from the HIJING model for two ranges of $|\Delta \eta|$ (middle row), and $v_2^{\text {est}} / v_2^{\text {sub}}$ defined in Eq.~\ref{eq:cl5} displayed as a function of $\pT$ in central \heau\  (left column), \dau\ (middle column), and \pau\ (right column) collisions. Only statistical uncertainties are shown.}\label{fig:12}
\end{figure*}

\begin{figure*}[htbp]
\centering
\includegraphics[width=0.85\linewidth]{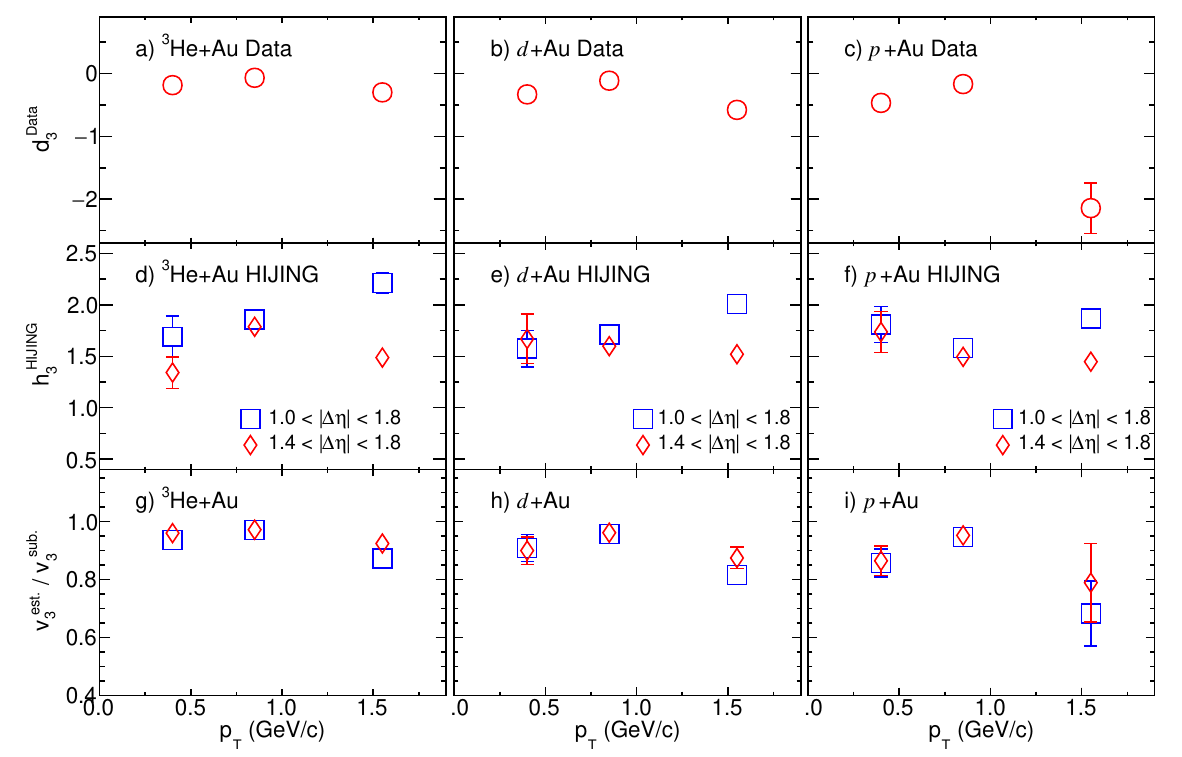}
%\caption{The values of $d_3 \equiv \frac{c_3^{pp}}{c_3} \frac{c_1}{c_1^{pp}}$ calculated from real data with $|\Delta \eta|>1$ (top row), $h_3 \equiv \frac{c_3^{pp \mathrm{hij}}}{c_3^{\mathrm{hij}}} \frac{c_1}{c_1^{pp}}$ derived from the HIJING model for two ranges of $|\Delta \eta|$ (middle row), and $v_3^{\text {est}} / v_3^{\text {sub}}$ defined in Eq.~\ref{eq:cl5} displayed as a function of $\pT$ in central \heau\  (left column), \dau\ (middle column), and \pau\ (right column) collisions. Only statistical uncertainties are shown.}\label{fig:13}
\caption{The values of $d_3$ defined in Eq.~\ref{eq:cl4} and calculated from real data with $|\Delta \eta|>1$ (top row), $h_3$ defined in Eq.~\ref{eq:cl2} and derived from the HIJING model for two ranges of $|\Delta \eta|$ (middle row), and $v_3^{\text {est}} / v_3^{\text {sub}}$ defined in Eq.~\ref{eq:cl5} displayed as a function of $\pT$ in central \heau\  (left column), \dau\ (middle column), and \pau\ (right column) collisions. Only statistical uncertainties are shown.}\label{fig:13}
\end{figure*}

\subsection{Closure test of the non-flow subtraction with HIJING}\label{sec:3.5}
In this section, we present a closure test of the non-flow subtraction method with the HIJING model. This test aims to assess the validity of the non-flow subtraction procedures by comparing the results obtained from experimental data with those from the HIJING model, which only includes non-flow correlations.

As discussed in the previous section, various non-flow subtraction methods differ mainly in estimating the scale factor $K$ to be multiplied to the \pp\ Fourier harmonics, 
\begin{align}\label{eq:cl1}
c^{\mathrm {sub}}_{n} = c_{n} - K\times{c^{pp}_{n}}\;,
\end{align}
where $K$ is equal to $c_{0}^{pp}/c_{0}$ for the $c_0$ method and $c_{1}/c_{1}^{pp}$ for the $c_1$ method. However, for the following discussion, we will focus on the default $c_1$ method, for which $K= c_{1}/c_{1}^{pp}$.

{\bf Method}. Residual non-flow can be estimated directly using models like HIJING~\cite{CMS:2010ifv,Lim:2019cys}. However, this approach relies on the model accurately reproducing the main features of jet-like correlations in $p+p$ collisions, such as their $\Dphi$, $\Deta$, and $\pT$ dependence, which is not the case. Instead, our approach takes non-flow features directly from $p+p$ data, using the difference in the factor $K$ between the HIJING model and data for the closure test. This procedure relies only on the HIJING model to estimate the scaling behavior of non-flow as a function of $\nch$ and between different collision systems, not its absolute yield.

%where $h_n$ can be calculated by scaling the Fourier harmonics in \pp\ collisions to match those in $p/d/^{3}$He+Au collisions:

{\bf Scaling factors}. The factor $K= c_{1}/c_{1}^{pp}$ in Eq.~\ref{eq:cl1} may be overestimated or underestimated by a factor $h_n$ that depends on the harmonic number $n$. While $h_n$ cannot be directly determined from experimental data, it can be explored using the HIJING model:
\begin{align}\label{eq:cl2}
c_{n}^{\mathrm {sub,hij}} =c_{n}^{\mathrm {hij}} - \frac{K}{h_n}\times{c^{pp,\mathrm {hij}}_{n}}=0\rightarrow h_n=K\frac{c^{pp,\mathrm {hij}}_{n}}{c_{n}^{\mathrm {hij}}}\;.
\end{align}
Here, $c^{pp,\mathrm {hij}}_{n}$ and $c_{n}^{\mathrm {hij}}$ are the corresponding Fourier harmonics in HIJING simulations. Notably, $h_n$ is always positive as both $c^{pp}_{n}$ and $c_{n}$ have the same sign in the HIJING model.

We consider two scenarios for $h_n$ with respect to its harmonic number,
\begin{itemize}
\item   For $n=2$,  $c^{pp}_{2}$ in Eq.~\ref{eq:cl1} is positive, so $h_2>1$ ($h_2<1$) indicates overestimation (underestimation) of non-flow contributions for elliptic flow measurements.
\item   For $n=3$,  $c^{pp}_{3}<0$, so $h_3<1$ ($h_3>1$) implies overestimation (underestimation) of non-flow contributions for triangular flow measurements.
\end{itemize}
These scenarios lead to different impacts of non-flow subtractions on $v_2$ and $v_3$ in the context of the HIJING model.

{\bf Assessing non-flow closure}. The degree to which the $c_1$ method accurately characterizes non-flow correlations can be assessed using:
\begin{align}\label{eq:cl3}
\frac{c^{\mathrm {est}}_{n}}{c^{\mathrm {sub}}_{n}} = \frac{c_{n} -  (K/h_n)\times{c^{pp}_{n}}}{c_{n} - K\times{c^{pp}_{n}}} = \frac{1-d_n/h_n}{1-d_n}\;,
\end{align}
where $c^{\mathrm {est}}_{n}$ represents the two-particle flow coefficients calculated using the scale factor from HIJING. This value deviates from $c^{\mathrm {sub}}_{n}$ if $h_n\neq1$. Additionally, we define a new quantity $d_n$ from real data,
\begin{align}\label{eq:cl4}
d_n=K\frac{c^{pp}_{n}}{c_{n}}\;.
\end{align}
Its form is similar to $h_n$ in Eq.~\ref{eq:cl2}, though with different behavior in terms of its sign:
\begin{itemize}
\item $d_2$ is always positive since both $c^{pp}_{2}$ and $c_2$ in the data are positive.
\item $d_3$ is always negative since $c^{pp}_{3}<0$ and $c_3>0$ in the data. 
\end{itemize}

%i.e. flow at low $\pT$, covered by associated particles, is insensitive to the non-flow subtraction procedure,
This distinction leads to a redefinition of Eq.~\ref{eq:cl3} for the two harmonics, providing an estimate of the potential change in $v_n$ due to non-flow subtraction uncertainties,
\begin{align}\label{eq:cl5}
&\frac{v^{\mathrm {est}}_{2}}{v^{\mathrm {sub}}_{2}} \approx \frac{c^{\mathrm {est}}_{2}}{c^{\mathrm {sub}}_{2}} = \frac{1-|d_2|/h_2}{1-|d_2|},\\\label{eq:cl6}
&\frac{v^{\mathrm {est}}_{3}}{v^{\mathrm {sub}}_{3}} \approx \frac{c^{\mathrm {est}}_{3}}{c^{\mathrm {sub}}_{3}} =  \frac{1+|d_3|/h_3}{1+|d_3|},
\end{align}
where we have used the factorization assumption and the observation that $v^{\mathrm {est a}}_{2}/v^{\mathrm {sub,a}}_{2}\approx1$, 
\begin{align}\label{eq:cl7}
\frac{c^{\mathrm {est}}_{n}}{c^{\mathrm {sub}}_{n}} = \frac{v^{\mathrm {est t}}_{n}}{v^{\mathrm {sub,t}}_{n}} \frac{v^{\mathrm {est a}}_{n}}{v^{\mathrm {sub,a}}_{n}}\approx  \frac{v^{\mathrm {est t}}_{n}}{v^{\mathrm {sub,t}}_{n}}
\end{align}

Given the differing signs between Eq.~\ref{eq:cl5} and Eq.~\ref{eq:cl6}, it is expected that $v_3^{\text {est}}/v_3^{\text {sub}}$ would be closer to unity than $v_2^{\text {est}}/v_2^{\text {sub}}$ for the same $h_n$ and $|d_n|$ values. Thus, $v_3$ is more robust against non-flow subtraction uncertainties than $v_2$.

{\bf Results}. In Fig.~\ref{fig:12}, the values of $d_2$ from data, $h_2$ from HIJING, and the resulting $v_2^{\text {est}} / v_2^{\text {sub}}$ are shown as functions of $\pT$ for the three systems. The top row presents the calculated $d_2$ using $|\Delta \eta|>1$. The increase in $d_2$ with $\pT$ reflects larger non-flow contributions from the away-side jet. In \pau\ collisions, $d_2$ reaches 0.6--0.8 at high $\pT$, indicating an enhanced sensitivity to the systematic uncertainties of non-flow subtraction.
%significant reduction in the denominator of Eq.~\ref{eq:cl5} and an 

The middle row shows $h_2$ from HIJING as a function of $\pT$ for the three systems. The simulation indicates broader near-side peaks in HIJING compared to the data (as seen in Fig.~\ref{fig:app6} in Appendix~\ref{sec:app}). Consequently, even with a $|\Delta \eta|>1$ cut, the residual near-side jet in the HIJING model may still bias the estimated $h_2$ value more than in data. Applying a stricter cut of $|\Delta \eta|>1.4$ results in correlation functions more similar to data. However, $h_2$ values are fortuitously similar for both cuts, always above unity and increasing with $\pT$.

%Nevertheless, we calculate $h_2$ from both $|\Deta|>1$ and $|\Deta|>1.4$ cuts, whose values are fortuitously similar. The values of $h_2$ are always above unity: they increase with $\pT$, but are quite similar in the three systems. 

The bottom row of Fig.~\ref{fig:12} shows the results of $v_2^{\text {est}} / v_2^{\text {sub}}$. With $h_2>0$, non-flow scale factors obtained from HIJING are smaller than those derived from data, resulting in larger $v_2$ values. If scale factors from HIJING are accurate, these results suggest that the $c_1$ method overcorrects $v_2$ in data. Overcorrection amounts to approximately 0--8\% in \heau, 0--15\% in \dau, and 0--50\% in \pau\ collisions across the measured $\pT$ range.

In Fig.~\ref{fig:13}, results of $d_3$ from data and $h_3$ from HIJING, and the corresponding $v_3^{\text {est}} / v_3^{\text {sub}}$ are shown as functions of $\pT$ for the three collision systems. The top row shows the calculated $d_3$ values, consistently negative as expected, but with increasing magnitude at higher $\pT$.

The middle row shows $h_3$ from HIJING as a function of $\pT$ for the three systems. $h_3$ values are above unity, ranging from around 1.5--2.0, with weak dependence on $\pT$. This finding implies that the numerator in Eq.~\ref{eq:cl6} is smaller than the denominator, indicating that $v_3^{\text {est}}<v_3^{\text {sub}}$. Results in the bottom row confirmed this, showing that $v_3^{\text {est}}$ is smaller than $v_3^{\text {sub}}$ by approximately 5--10 \% in \heau, 10--15\% in \dau, and 15--20 \% in \pau\ collisions. This suggests that the $c_1$ method overestimates the $v_3$ signal by these amounts, assuming HIJING accurately describes non-flow correlations.

To sum up, the scaling behavior of non-flow in the HIJING model shows some differences from real data. Using HIJING scale factors to adjust non-flow subtraction, $v_2$ values remain largely consistent except in \pau\ collisions at high $\pT$. Conversely, $v_3$ values would be slightly reduced by 4--25\% across all collision systems and $\pT$ ranges.

A previous study in Ref.\cite{Lim:2019cys} explored the performance of non-flow subtraction using HIJING. The study identified residual non-closure of the subtraction, although it was conducted within a somewhat different $\pT$ range. However, based on STAR's analysis method and kinematic selection, the impact of this non-closure on $v_3$ results is modest and within experimental systematic uncertainties (see Table~\ref{tab:3}).

%A previous study in Ref.~\cite{Lim:2019cys} explored the performance of the nonflow subtraction procedure using the HIJING model. The study identified residual non-closure of the subtraction, although it was conducted within a somewhat different $\pT$ range. The findings indicated that the non-closure effect is significant in \pau\ collisions at high $\pT$ ($>1$ GeV/$c$), which aligns with the observations made in this analysis. 
%It is important to note that, however, based on the analysis method and kinematic selection employed by STAR, the impact of this non-closure has only a modest effect on the $v_3$ results, and is well within the experimental systematic uncertainties (see Table~\ref{tab:3}).

\subsection{Dependence on the $\Delta\eta$ selection}\label{sec:3.3} 
This section examines the impact of varying the pseudorapidity gap $(|\Deta|)$ between particle pairs, aiming to further evaluate the robustness of non-flow subtraction methods. The default gap selection of $|\Deta|>1.0$ is chosen to effectively suppress near-side non-flow correlations and reduce the influence of away-side non-flow effects. Our analysis centers on the default $c_1$ method, and we explore how varying the $|\Deta|$ cut influences the stability of the extracted $v_n$ values.

\begin{figure}[ht]
\includegraphics[width=1.\linewidth]{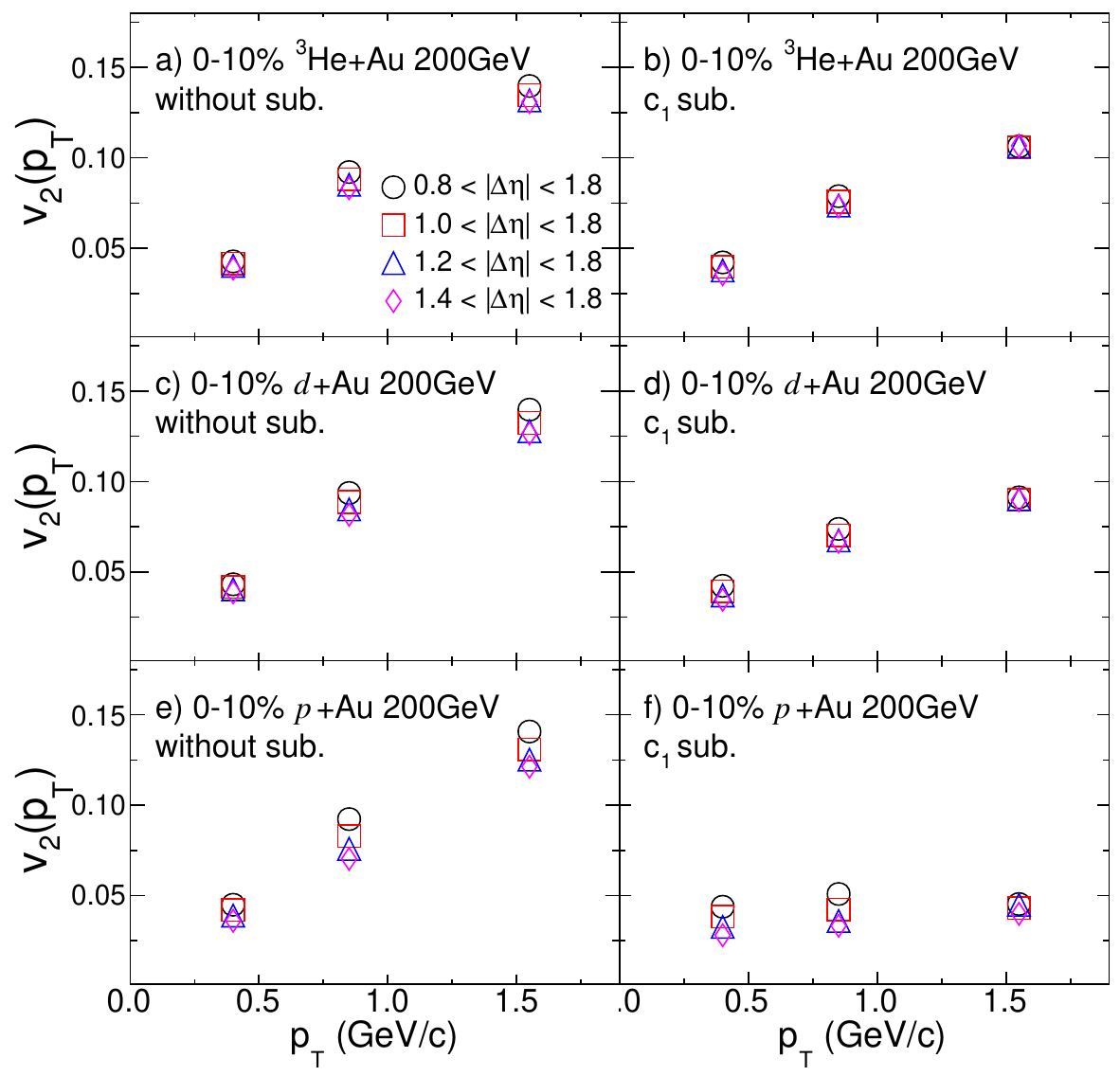}
\caption{Values of $v_2$ obtained from different $\Deta$ selections: $|\Delta\eta|>$~0.8, 1.0, 1.2 and 1.4 in top 0--10\% \pau, \dau\ and \heau\ collisions at \TopE. The left column shows results before non-flow subtraction, and the right column shows results after applying the $c_1$ subtraction method. Only statistical uncertainties are shown.}
\label{fig:9}
\end{figure}

\begin{figure}[ht]
\includegraphics[width=1.\linewidth]{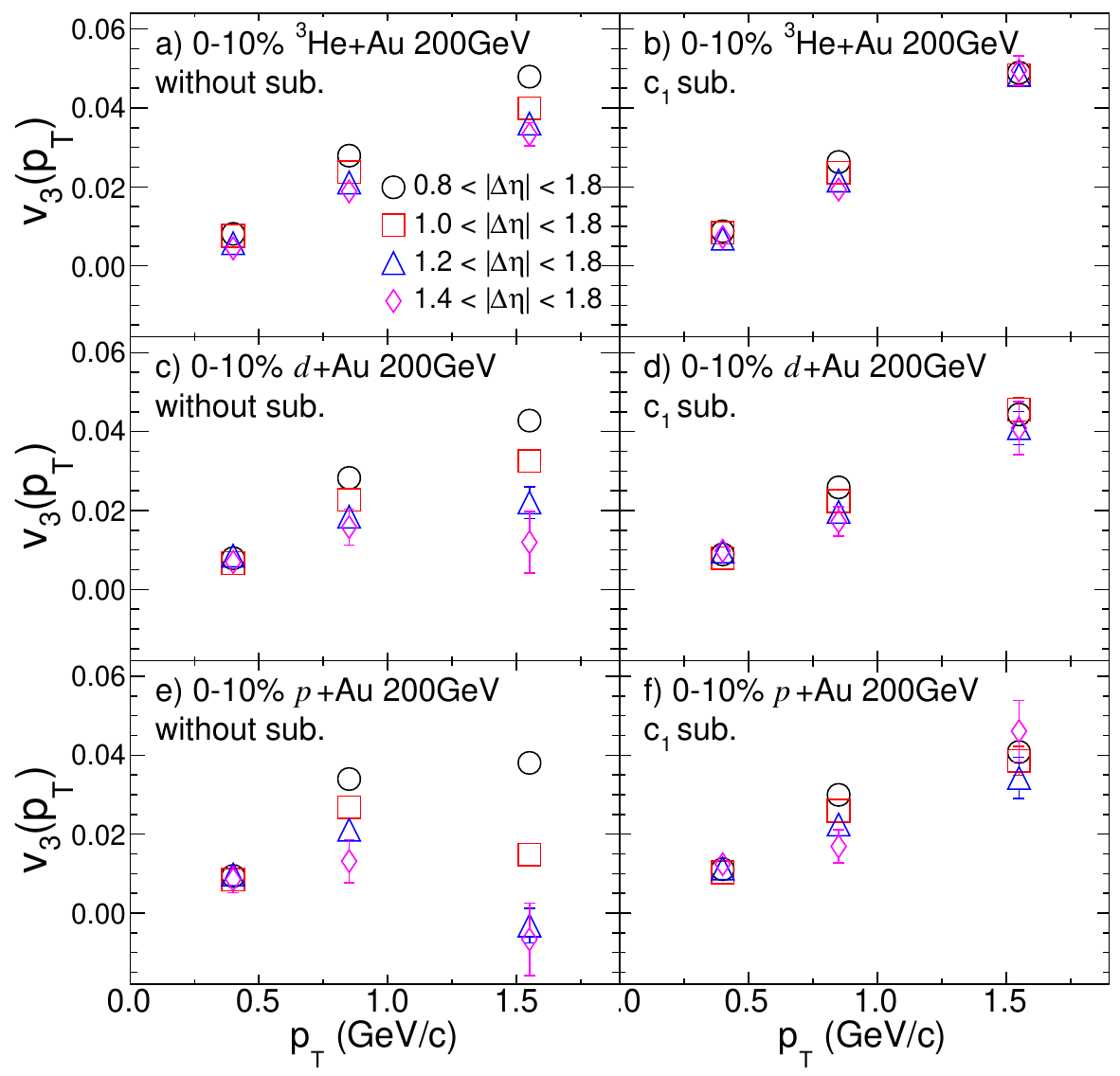}
\caption{Values of $v_3$ obtained from different $\Deta$ selections: $|\Delta\eta|>$~0.8, 1.0, 1.2 and 1.4 in top 0--10\% \pau, \dau\ and \heau\ collisions at \TopE. The left column shows results before non-flow subtraction, and the right column shows results after applying the $c_1$ subtraction method. Only statistical uncertainties are shown.}
\label{fig:10}
\end{figure}

We systematically vary the $|\Deta|$ cut and analyze the resulting $v_2$ and $v_3$ values, both with and without non-flow subtraction. The results for $v_2$ are presented in Fig.~\ref{fig:9}, and those for $v_3$ in Fig.~\ref{fig:10}. This approach reveals key insights into the influence of non-flow correlations.

The primary source of non-flow in $v_2$ measurements arises from away-side jet-like correlations. Increasing the $|\Deta|$ cut from $|\Deta|>0.8$ to $|\Deta|>1.4$ only slightly further suppresses residual near-side non-flow. This accounts for the independence of $v_2$ on $|\Deta|$ before non-flow subtraction in the left column of Fig.~\ref{fig:9}. After non-flow subtraction via the $c_1$ method, the resulting $v_2$ values are significantly lower yet remain nearly independent of the $|\Deta|$ cut.

% methodology effectively eliminates most non-flow correlations, as shown in the right column of Fig.~\ref{fig:9}, yielding $v_2$ values that are smaller yet remain nearly independent of the $|\Deta|$ cut.
%remove , though it results in only a slight reduction in overall non-flow contribution, however, experiences only a slight reduction with increasing $|\Deta|$.

The behavior of $v_3$ is more complex. As previously mentioned, residual near-side jet correlations contribute positively to $v_3$, while away-side jet-like correlation tends to reduce the $v_3$. This means that non-flow contributions from near- and away-side jets compete and can partly cancel each other out. This interplay is evident in the left column of Fig.~\ref{fig:10}, where increasing the $|\Deta|$ cut reduces the positive contribution from the near-side jet, leading to a decrease in the extracted $v_3$. This trend is observed across all three collision systems, with the most significant impact seen in \pau\ collisions at high $\pT$ (bottom-left panel of Fig.~\ref{fig:10}). The variation follows the order \pau$<$\dau$<$\heau, simply because larger system has higher multiplicity and is less affected by non-flow and multiplicity selection biases

However, after non-flow subtraction, the $v_3$ values obtained from various $|\Deta|$ cuts converge, as shown in the right column of Fig.~\ref{fig:10}. This alignment indicates that both near- and away-side non-flow contributions have been effectively eliminated. Interestingly, the $v_3$ results for $|\Deta|>0.8$ are nearly identical before and after non-flow subtraction, suggesting a fortuitous cancellation between the positive near-side and negative away-side jets contributions.

In conclusion, this investigation demonstrates that a $|\Deta|>1.0$ selection is optimal for the STAR TPC acceptance, balancing the suppression of non-flow effects with statistical precision in determining $v_2$ and $v_3$.

\subsection{Non-flow bias in selecting high-multiplicity events}\label{sec:3.4}
\begin{figure}[h!]
\centering
\includegraphics[width=1.\linewidth]{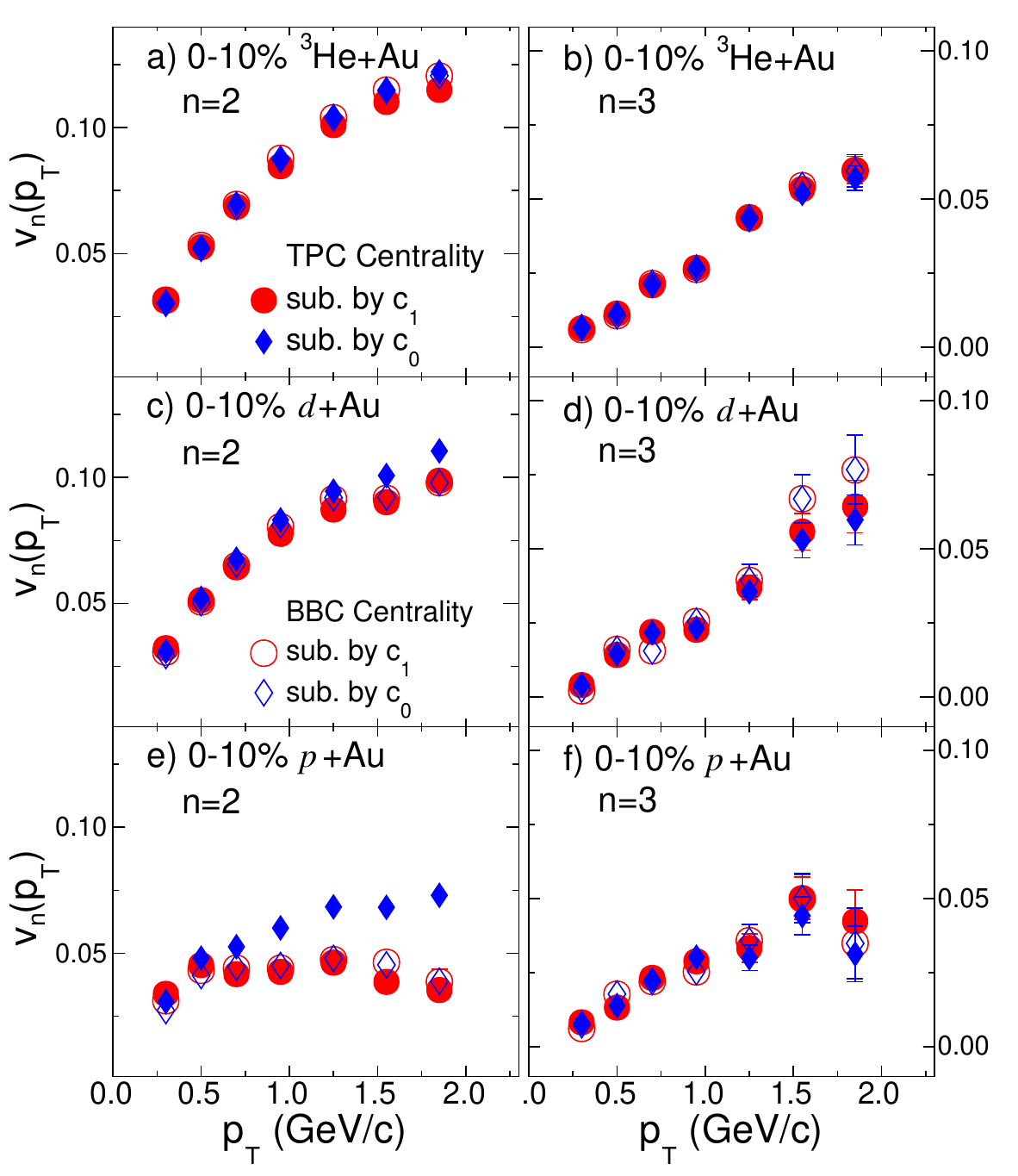}
\caption{Values of $v_2$ (left) and $v_3$ (right) in central collisions based on TPC selection (filled symbols) and BBC selection (open symbols), obtained using the $c_1$ method (circles) and $c_0$ method (diamond) in 0--10\% \heau\ (top row), 0--10\% \dau\ (middle row) and 0--10\% \pau\ (bottom row) collisions. These results are obtained with a requirement of $|\Delta\eta| > 1.0$. Only statistical uncertainties are displayed.}\label{fig:11}
\end{figure}

By default, centrality selection is based on the $N_{\mathrm{ch}}^{\mathrm{raw}}$ measured in the TPC as detailed in Sec.~\ref{sec:2.1b}. This approach allows us to reach high $N_{\mathrm{ch}}^{\mathrm{raw}}$ values for flow measurement. However, this approach may introduce potential biases in jet fragmentation, which could subsequently affect the non-flow contributions. To investigate these potential biases stemming from the choice of high-multiplicity events on non-flow correlations, we conducted an analysis using two distinct centrality definitions: one based on the TPC and the other on the forward rapidity multiplicity measured by $\Sigma Q_{\mathrm{BBCE}}$. A comparison of the $v_n$ values obtained using these two centrality definitions, with the $c_0$ and $c_1$ non-flow subtraction methods, is shown in Figure~\ref{fig:11}.

The results reveal a high level of consistency between the two non-flow subtraction methods when adopting the BBC-based centrality selection. However, when the TPC-based centrality selection is used, the $v_2$ values show significant discrepancies between the two methods, especially in \pau\ collisions at high $\pT$. These discrepancies can be attributed to biases induced by the away-side non-flow effect on the per-trigger yield, which is underestimated by the scaling factor $c_0^{pp} / c_0$ in the $c_0$ method (Eq.~\ref{eq:nf1}). In contrast, the scale factor $c_1 / c_1^{pp}$ employed in the $c_1$ method (Eq.~\ref{eq:nf3}) accurately reflects the magnitude of the away-side non-flow, regardless of the centrality definition. This comparison strongly suggests that the $c_1$ method is more reliable than the $c_0$ method for assessing the non-flow contribution.

For $v_3$, the results are significantly less sensitive to the choice of centrality definition. This outcome aligns with our previous findings that $v_3$ values are less affected by non-flow correlations under the kinematic criteria used in this analysis.

The multiplicity selection bias could in principle be further checked by the even smaller system such as the $p+$Al collisions in 2015. However, STAR did not record sufficient event statstics for such a study.
 
\begin{table*}[htbp!]
   \begin{center}
   \begin{tabular}{ c| c  c|  c  c  c | c c}
     \hline\hline
     & & & & & & &\\
     Sources& &$\pT$ range(GeV/$c$) & 0--10\% \heau & 0--10\% \dau &  0--10\% \pau  & 0--2\% \dau & 0--2\% \pau\\
     & & & & & & &\\
    \hline \hline
    \multirow{6}{*}{Track selection} & \multirow{3}{*}{$v_{2}$} & 0.2$<\pT<$0.6 & $<$ 2\% & $<$ 2\%  & $<$ 5\%  & $<$ 2\% & $<$ 2\%\\
    &&0.6$<\pT<$1.1& $<$ 2\% & $<$ 2\%  & $<$ 5\%  & $<$ 2\% & $<$ 2\%\\
    &&1.1$<\pT<$2.0& $<$ 2\% & $<$ 2\%  & $<$ 5\%  & $<$ 2\% & $<$ 2\%\\
    \cline{2-8}
    & \multirow{3}{*}{$v_{3}$} & 0.2$<\pT<$0.6 & $<$ 2\% & $<$ 6\%  & $<$ 4\% & $<$ 6\% & $<$ 10\%\\
    &&0.6$<\pT<$1.1& $<$ 2\% & $<$ 5\%  & $<$ 9\% &$<$ 5\% & $<$ 9\%\\
    &&1.1$<\pT<$2.0& $<$ 2\% & $<$ 4\%  & $<$ 3\% &$<$ 4\% & $<$ 3\%\\
     \hline
     \hline
    \multirow{6}{*}{Matching to TOF/HFT} & \multirow{3}{*}{$v_{2}$} & 0.2$<\pT<$0.6 & $<$ 2\% & $<$ 2\%  & $<$ 2\% & $<$ 2\% & $<$ 2\% \\
    &&0.6$<\pT<$1.1& $<$ 2\% & $<$ 2\%  & $<$ 3\% & $<$ 2\% & $<$ 2\% \\
    &&1.1$<\pT<$2.0& $<$ 2\% & $<$ 2\%  & $<$ 3\% & $<$ 2\% & $<$ 2\% \\
    \cline{2-8}
    & \multirow{3}{*}{$v_{3}$} & 0.2$<\pT<$0.6 & $<$ 3\% & $<$ 5\%  & $<$ 3\% & $<$ 8\% & $<$ 3\%\\
    &&0.6$<\pT<$1.1& $<$ 3\% & $<$ 3\%  & $<$ 3\%  & $<$ 2\% & $<$ 3\% \\
    &&1.1$<\pT<$2.0& $<$ 3\% & $<$ 8\%  & $<$ 12\% & $<$ 7\% & $<$ 5\% \\
    \hline
    \hline
  \multirow{2}{*}{Luminosity dependence}& $v_{2}$ & 0.2$<\pT<$2.0 & $<$ 2\% & $<$ 2\%  & $<$ 2\% & $<$ 2\% & $<$ 2\% \\
   
    \cline{2-8}
    & $v_{3}$ & 0.2$<\pT<$2.0 & $<$ 5\% & $<$ 5\%  & $<$ 5\% & $<$ 5\% & $<$ 5\%\\
    
    \hline
    \hline
    
    \multirow{6}{*}{Non-flow subtraction} & \multirow{3}{*}{$v_{2}$} 
     &0.2$<\pT<$0.6& $<$ 13\%& $<$ 15\%  & $<$ 28\% & $<$ 15\% & $<$ 29\% \\
    &&0.6$<\pT<$1.1& $<$ 8\% & $<$ 11\%  & $<$ 34\% & $<$ 9\%  & $<$ 16\%  \\
    &&1.1$<\pT<$2.0& $<$ 9\% & $<$ 12\%  & $<$ 64\% & $<$ 10\% & $<$ 24\% \\
    \cline{2-8}
    & \multirow{3}{*}{$v_{3}$} 
     &0.2$<\pT<$0.6& $<$18\% & $<$ 21\%  & $<$ 29\% & $<$6\% & $<$ 27\%  \\
    &&0.6$<\pT<$1.1& $<$ 17\%& $<$ 21\%  & $<$ 34\% & $<$ 12\% & $<$ 26\% \\
    &&1.1$<\pT<$2.0& $<$ 8\% & $<$ 12\%  & $<$ 24\% & $<$ 17\% & $<$ 13\% \\
    \hline
     \hline
     
     \multirow{6}{*}{Total} & \multirow{3}{*}{$v_{2}$} 
     &0.2$<\pT<$0.6& $<$ 13\%& $<$ 16\%  & $<$ 29\% & $<$ 16\% & $<$ 25\%   \\
    &&0.6$<\pT<$1.1& $<$ 9\% & $<$ 12\%  & $<$ 34\% & $<$  9\% & $<$ 16\%  \\
    &&1.1$<\pT<$2.0& $<$ 9\% & $<$ 13\%  & $<$ 65\% & $<$  10\% & $<$ 24\% \\
    \cline{2-8}
    & \multirow{3}{*}{$v_{3}$} 
     &0.2$<\pT<$0.6& $<$ 19\% & $<$ 21\%  & $<$ 30\% & $<$ 13\% & $<$ 29\% \\
    &&0.6$<\pT<$1.1& $<$ 19\% & $<$ 22\%  & $<$ 34\% & $<$ 14\% & $<$ 28\% \\
    &&1.1$<\pT<$2.0& $<$ 11\% & $<$ 13\%  & $<$ 28\% & $<$ 19\% & $<$ 15\% \\
    \hline
     \hline
    \end{tabular}
   \caption{\label{tab:3} Main sources of systematic uncertainties for $v_2$ and $v_3$ measurements in 0--10\% central \heau, \dau\ and \pau\ collisions and 0--2\% ultracentral \dau\ and \pau\ collisions.}
   \end{center}
 \end{table*}

\section{Systematic Uncertainties}\label{sec:4}
The systematic uncertainties in $v_n$ measurements arise from track selection criteria, background tracks, residual pileup events, and non-flow subtraction procedures. For each source of uncertainties, the entire analysis pipeline, including non-flow subtraction, is repeated, and the resulting deviations from the default results are reported as uncertainties.

The impact of track selection is assessed by varying the TPC hit selection from 16 to 25 hits and varying the DCA cut. These variations result in changes of less than 5\% for $v_2$ and less than 10\% for $v_3$ across all three collision systems. Additionally, the criteria for matching tracks to fast detectors, crucial for background track elimination, are modified by requiring track matching to only TOF or either TOF or HFT. This adjustment induces deviations of less than 2\% for $v_2$ and under 5\% for $v_3$ in \heau\ and \dau\ collisions. In \pau\ collisions, the variation ranges from 2\% to 7\% for $v_2$ and remains under 5\% for $v_3$.

Differences in luminosity conditions across \pp\ and $p/d/^{3}$He+Au collisions can slightly vary track reconstruction efficiency. To account for this, luminosity-dependent scaling factors are incorporated into the two-particle correlation analysis. To evaluate the effect of luminosity fluctuations on track quality and background contamination, the data for each collision system is divided into subsets based on average luminosities as measured by the STAR BBC. Correlation analyses performed on these subsets reveal minimal dependence on luminosity condition, resulting in a 2\% uncertainty for $v_2$ and a 5\% uncertainty for $v_3$ across all systems.

The largest source of systematic uncertainty arises from the limited understanding of non-flow contributions. Section~\ref{sec:3} provides comprehensive discussions of non-flow subtraction methods and their effectiveness. To quantify this uncertainty, $v_n$ values are compared across four different subtraction methods and four distinct  $\Delta\eta$ gaps ($|\Delta\eta|>$0.8, 1.0, 1.2, and 1.4). Additionally, comparisons are made between correlations involving same-charge pairs and opposite-charge pairs to assess the impact of residual contributions from near-side jet fragmentation.

The default results are obtained using the $c_1$ method with $|\Delta\eta|>$1.0. The largest deviation from the other three subtraction methods is designated as the systematic uncertainty associated with non-flow subtraction. These uncertainties are then combined with variations among different $\Delta\eta$ gaps and those between same-charge and opposite-charge correlations. The overall uncertainty is below 15\% (21\%) for $v_2$($v_3$)  in \heau\ and \dau\ collisions. However, in \pau\ collisions, the uncertainty is significantly higher, reaching 65\% for $v_2$ and 35\% for $v_3$ at high $\pT$. Notably, the uncertainties related to non-flow subtraction methods are substantially smaller for the most central 0--2\% \pau\ collisions compared to those for the 0--10\% \pau\ collisions.

The uncertainties originating from the aforementioned sources are combined in quadrature, with non-flow subtraction being the dominant contributor. A detailed breakdown of systematic uncertainties is provided in Table~\ref{tab:3}.

\begin{figure*}[htbp]
\centering
\includegraphics[width=0.85\linewidth]{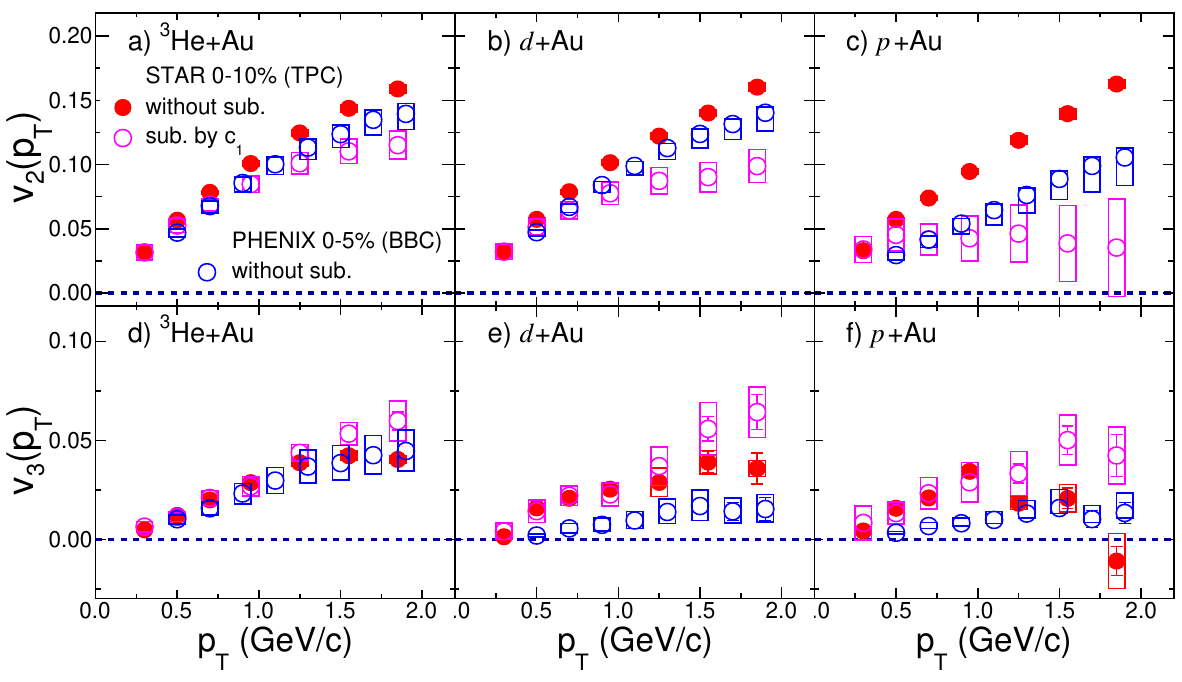}
\caption{Comparison of the $v_2(\pT)$ (top row) and $v_3(\pT)$ (bottom row) between measurements obtained by PHENIX (open blue circles), STAR results without non-flow subtraction (solid red circles), and STAR results with non-flow subtraction based on $c_1$ method (pink open circles). The boxes indicate the systematic uncertainties. The PHENIX results are for 0--5\% centrality, while the STAR results are for 0--10\% centrality. Systematic uncertainties for STAR results are shown for without subtraction and with the $c_1$ subtraction method.}
\label{fig:14}
\end{figure*}
\begin{figure*}[htbp]
\begin{center}
  \includegraphics[width=0.85\linewidth]{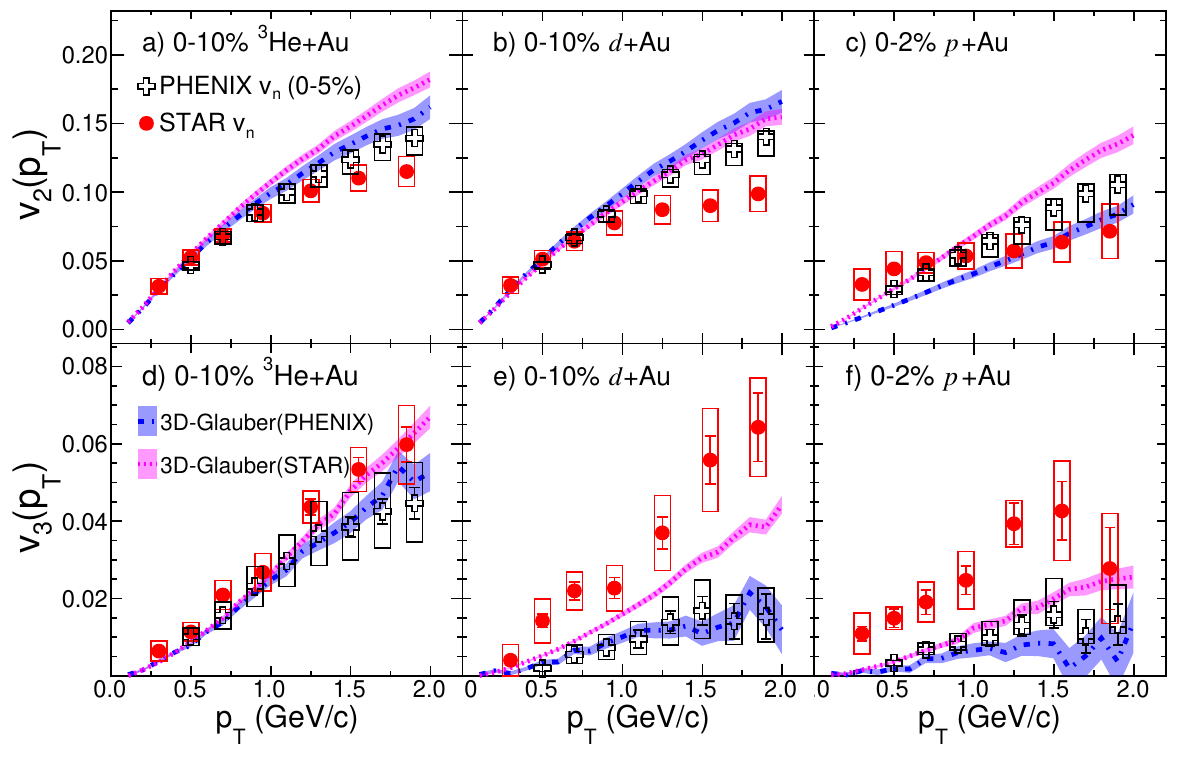}
  \caption{Comparison between the STAR results based on the $c_1$ method and PHENIX measurements, along with 3D-Glauber calculations using middle-middle (STAR) or middle-backward (PHENIX) correlations. This comparison highlights a significant difference in $v_3(\pT)$ between the two experiments, partially attributed to flow decorrelations according to the calculations~\cite{Zhao:2022ugy}. The boxes indicate the systematic uncertainties. For \pau\ collisions, STAR results are shown for 0--2\% centrality instead of the 0--10\% as in Fig.~\ref{fig:14}.}
  \label{fig:15}
  \end{center}
\end{figure*}

\section{Results and discussions}\label{sec:5}

\subsection{Comparison with previous results and model predictions}\label{sec:5.1}
The flow results from the STAR and PHENIX experiments in small collision systems show some discrepancies, which can be attributed to various factors, including differences in kinematic selection, analysis techniques, residual non-flow correlations, and longitudinal dynamics. Understanding these differences is crucial for a proper interpretation of the data.

PHENIX measurements are obtained from multiple pairs of correlations involving different combinations of particles at midrapidity and in the backward Au-going direction: $|\eta^{a}| <$ 0.35, \mbox{-3.0 $< \eta^{b} <$ -1.0}, and \mbox{-3.9 $< \eta^{c} <$ -3.1}. Using the notation of flow vectors in a subevent as ${\bm Q}_n \equiv q_n e^{in\psi_n}$, the $v_n(\pT)$ at midrapidity ($|\eta^{a}| <$ 0.35) is computed using an event-plane method that assumes factorization among the pairs from any two subevents,
\begin{align}\label{eq:cp1a}
v_{n}^a(\pT) \approx \frac{\lr{q_{n}^a(\pT)\cos n(\psi_{n}^a(\pT)-\psi_{n}^c)}\sqrt{\lr{\cos n(\psi_{n}^a-\psi_{n}^b)}}}{\sqrt{\lr{\cos n(\psi_{n}^a-\psi_{n}^c)}\lr{\cos n(\psi_{n}^c-\psi_{n}^b)}}}\;,
\end{align}
where $q_{n}^a(\pT)$ and $\psi_{n}^a(\pT)$ denote the magnitude and direction (or event plane) of the flow vector ${\bm Q}_n(\pT)$ at midrapidity. The $\psi_{n}^c$ and $\psi_{n}^b$ are event plane angles calculated using all particles (without $\pT$ selection) within the acceptance of a specific subevent. 

In the low event plane resolution limit, this equation simplifies to the scalar product method result:
\small{\begin{align}\nonumber
v_{n}^a(\pT) &=\!\frac{\lr{q_{n}^a(\pT\!)q_{n}^c\!\cos n(\!\psi_{n}^a(\pT\!)\!-\!\psi_{n}^c)}\!\sqrt{\lr{q_{n}^aq_{n}^b\cos n\!(\!\psi_{n}^a\!-\!\psi_{n}^b)}}}{\sqrt{\lr{q_{n}^aq_{n}^c\cos n(\psi_{n}^a-\psi_{n}^c)}\lr{q_{n}^cq_{n}^b\cos n(\psi_{n}^c-\psi_{n}^b)}}}\\\label{eq:cp1b}
&=\frac{\lr{{\bm Q}_{n}^a(\pT){\bm Q}_{n}^{c*}}\sqrt{\lr{{\bm Q}_{n}^a{\bm Q}_{n}^{b*}}}} {\sqrt{\lr{{\bm Q}_{n}^a{\bm Q}_{n}^{c*}}\lr{{\bm Q}_{n}^c{\bm Q}_{n}^{b*}}}}\\\label{eq:cp1c}
&\equiv\frac{c_n(a(\pT),c)\sqrt{{c_n(a,b)}}} {\sqrt{{c_n(a,c)}{c_n(c,b)}}}\;.
\end{align}}\normalsize
where, for instance, $c_n(c,b)$ represents the two-particle correlation for all particles accepted in subevents ``$c$'' and ``$b$''. The $a(\pT)$ denotes that particles in subevent ``$a$'' are chosen from a certain $\pT$ range. 

In addition to Eq.~\ref{eq:cp1c}, two independent combinations can also be used to calculate $v_{n}^a(\pT)$,
\begin{align}\nonumber
v_{n}^a(\pT)&=\sqrt{\frac{c_n(a(\pT),c){c_n(a(\pT),b)}} {c_n(c,b)}}\\
v_{n}^a(\pT)&=\frac{c_n(a(\pT),b)\sqrt{{c_n(a,c)}}} {\sqrt{{c_n(a,b)}{c_n(c,b)}}}\;.
\end{align}
Assuming factorization relations such as ${c_n(a(\pT),c) =v_n^a(\pT) v_n^c}$ and $c_n(c,b)=v_n^c v_n^b$, as often done in experimental measurements, all three combinations reduce to the same $v_{n}^a(\pT)$. Such factorization relations are explicitly broken by residual non-flow effects~\cite{Liu:2020ely} and longitudinal decorrelations~\cite{Bozek:2015bna}. However, if these contributions are negligible, all three approaches are expected to yield equivalent results.

In contrast, STAR measurements are derived from correlations between particles in the same mid-rapidity interval $|\eta^{a,b}| <$ 0.9 but with a definite pseudorapidity gap $|\Deta|>1.0$ between the pairs, as defined in Eq.~\ref{eq:nf7}. This small pseudorapidity gap reduces the impact of longitudinal flow decorrelations, which could be significant in smaller \pau\ collisions~\cite{CMS:2015xmx}.

In the PHENIX measurement, non-flow contributions are not subtracted from each of the $c_n$ terms in Eq.~\ref{eq:cp1c}. Instead, these non-flow contributions are estimated using an approach similar to the $c_0$ method and are incorporated as asymmetric systematic uncertainties. However, we have demonstrated that the $c_0$ method, at least within the STAR acceptance, could underestimate non-flow and is also influenced by auto-correlation effects (Fig.~\ref{fig:7} and Fig.~\ref{fig:11}). Therefore, the $c_1$ method is considered closer to the true flow value.

Figure~\ref{fig:14} compares $v_2$ and $v_3$ results between the two experiments for similar $\pT$ and centrality ranges. The STAR results, based on the $c_1$ method, are presented. The $v_2$ results without non-flow subtraction in \heau\ and \dau\ collisions are slightly higher than those of PHENIX, but they are 60\% larger in \pau\ collisions. This discrepancy likely reflects greater non-flow contributions in the STAR measurements due to its smaller $\Deta$ gap and larger away-side non-flow contributions. After non-flow subtraction, aside from minor $\pT$-dependent differences for $\pT > 1$ GeV/$c$, where STAR results are systematically lower, the $v_2$ results are consistent between the two experiments within uncertainties.

Since non-flow estimates based on the $c_0$ method are accounted for as asymmetric systematic uncertainties in the PHENIX results, comparing them with STAR results obtained using the same method is useful. Figure~\ref{fig:app7} in the Appendix reveals that the STAR $v_2$ values from the $c_0$ method lie just below the lower limit of the uncertainty bands of the PHENIX results. In contrast, the STAR $v_3$ values from the $c_0$ method fall outside the uncertainty region of corresponding PHENIX results. This discrepancy may result from the longitudinal decorrelations.

Recent calculations utilizing a 3D-Glauber model~\cite{Zhao:2022ugy}, as depicted in Fig.~\ref{fig:15}, suggest that more pronounced decorrelation effects in the PHENIX results contribute to about half of the difference in $v_3$ between the two experiments. However, this model underestimates $v_3$ measurements from both experiments in \pau\ collisions.

In addition to the differences in residual non-flow correlations and longitudinal decorrelations,  the measurements are also influenced by variations in modeling the initial collision geometry and early-time transverse dynamics, which are common to both experiments. These aspects are further elaborated below.

\begin{figure*}[htbp]
\begin{center}
\includegraphics[width=0.9\linewidth]{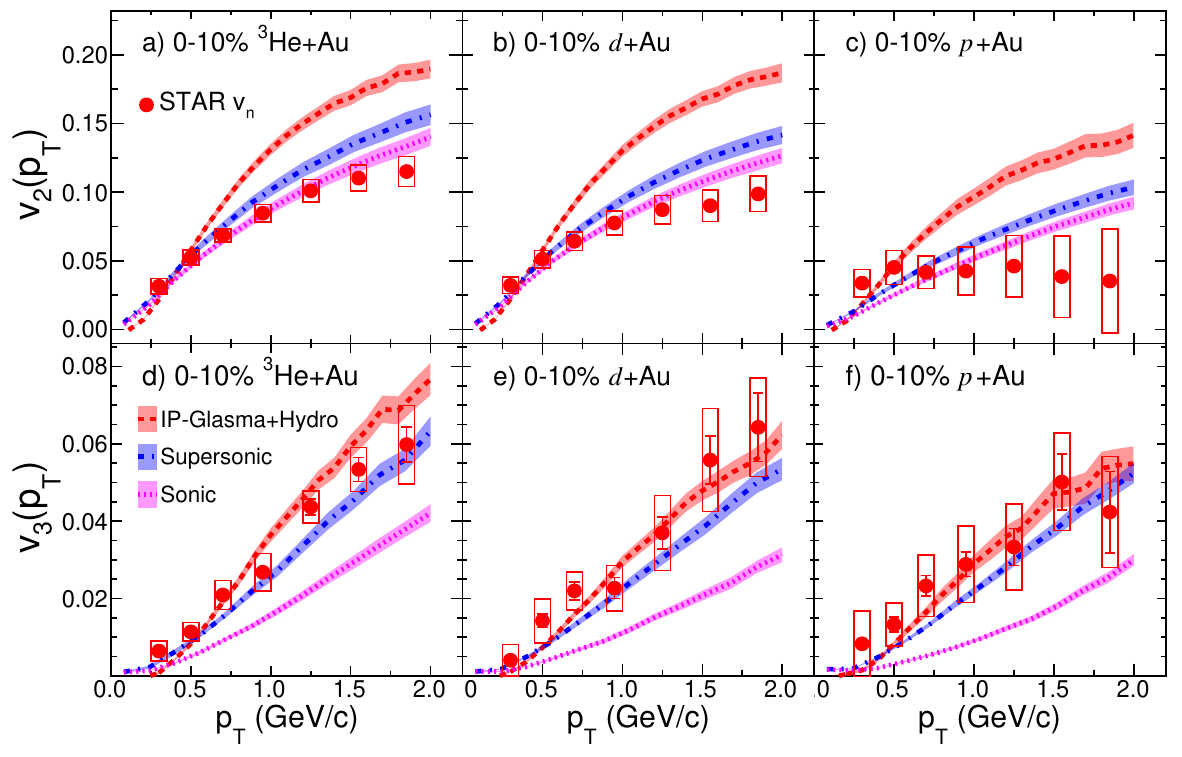}
\caption{Comparison of $v_2(\pT)$ (top row) and $v_3(\pT)$ (bottom row) values based on the $c_1$ method in 0--10\% most central \heau\ (left column), \dau\ (middle column), and \pau\ (right column) collisions with calculations from three hydrodynamical models: the {\sc sonic} model~\cite{Romatschke:2015gxa} (pink bands with dotted lines), the {\sc supersonic} model (blue bands with dash-dotted lines)~\cite{Romatschke:2015gxa} and IP-Glasma+Hydro model (red bands with dashed lines)~\cite{Schenke:2019pmk,Schenke:2020mbo}. The boxes are the systematic uncertainties.}
 
  \label{fig:16}
  \end{center}
\end{figure*}

Figure~\ref{fig:16} contrasts the $v_2$ and $v_3$ results from the three systems with three hydrodynamic model calculations that make distinct assumptions about the initial collision geometry and early dynamics. The {\sc sonic} model~\cite{Romatschke:2015gxa} incorporates viscous hydrodynamics with a nucleon Glauber initial geometry. The {\sc supersonic} model from the same reference introduces an additional pre-equilibrium flow phase, enhancing initial velocity fields during system evolution. The third model~\cite{Schenke:2019pmk,Schenke:2020mbo} combines IP-Glasma initial conditions with subnucleon fluctuations and pre-flow effects, MUSIC for hydrodynamic evolution, and UrQMD for hadronic phase interactions. All three models assume boost-invariant initial conditions, meaning that both non-flow and longitudinal dynamics are absent. The transport coefficients in these models, such as shear viscosity and freeze-out conditions, have been tuned to describe flow data in large Au+Au collision systems.

The comparison of these models with the experimental data yields interesting insights. The {\sc sonic} model underestimates the $v_3$ values observed across all three collision systems. The {\sc supersonic} model, which includes the pre-flow effect, achieves a better agreement with the experimental data. The IP-Glasma+hydro model successfully describes the $v_3$ results in all three systems but tends to overestimate the $v_2$ results. This comparison underscores the complexity of interpreting small system flow data. To truly understand the roles played by pre-equilibrium flow, nucleon fluctuations, and subnucleon fluctuations, comprehensive studies are necessary. These should include further model refinements, additional small system collision data, and more differential measurements.

Looking ahead, STAR has collected new \dau\ and $^{16}$O+$^{16}$O data in 2021 using the updated detector systems. These upgrades include the inner TPC, which extends tracking to $|\eta| <$ 1.5~\cite{Wang:2017mbk}, and the Event Plane Detector, capable of measuring charged particles in the 2.1 $< |\eta| <$ 5.3 range~\cite{Adams:2019fpo}. Utilizing this dataset will allow STAR to directly compare correlations obtained at midrapidity with those between the middle and backward regions. This comparison is expected to shed light on the roles of longitudinal decorrelation and non-flow correlations in small systems, and hence bridge the gap in our understanding between the STAR and PHENIX measurements. 

The symmetric $^{16}$O+$^{16}$O system, similar in size to \dau\ in terms of number of collided nucleons but markedly distinct geometry, is anticipated to be less influenced by subnucleon fluctuations and centrality selection biases. Comparing this system with existing small system data at RHIC holds the potential to disentangle various competing effects related to initial geometry and hydrodynamic evolution. Additionally, a comparison with future $^{16}$O+$^{16}$O data at the LHC, scheduled for collection in 2024, will provide direct insights into the energy dependence of pre-flow and longitudinal dynamics. These future endeavors are essential for a more comprehensive understanding of the intricate interplay between small system dynamics and the underlying physics mechanisms.

\begin{figure}[ht]
\centering
\includegraphics[width=1.\linewidth]{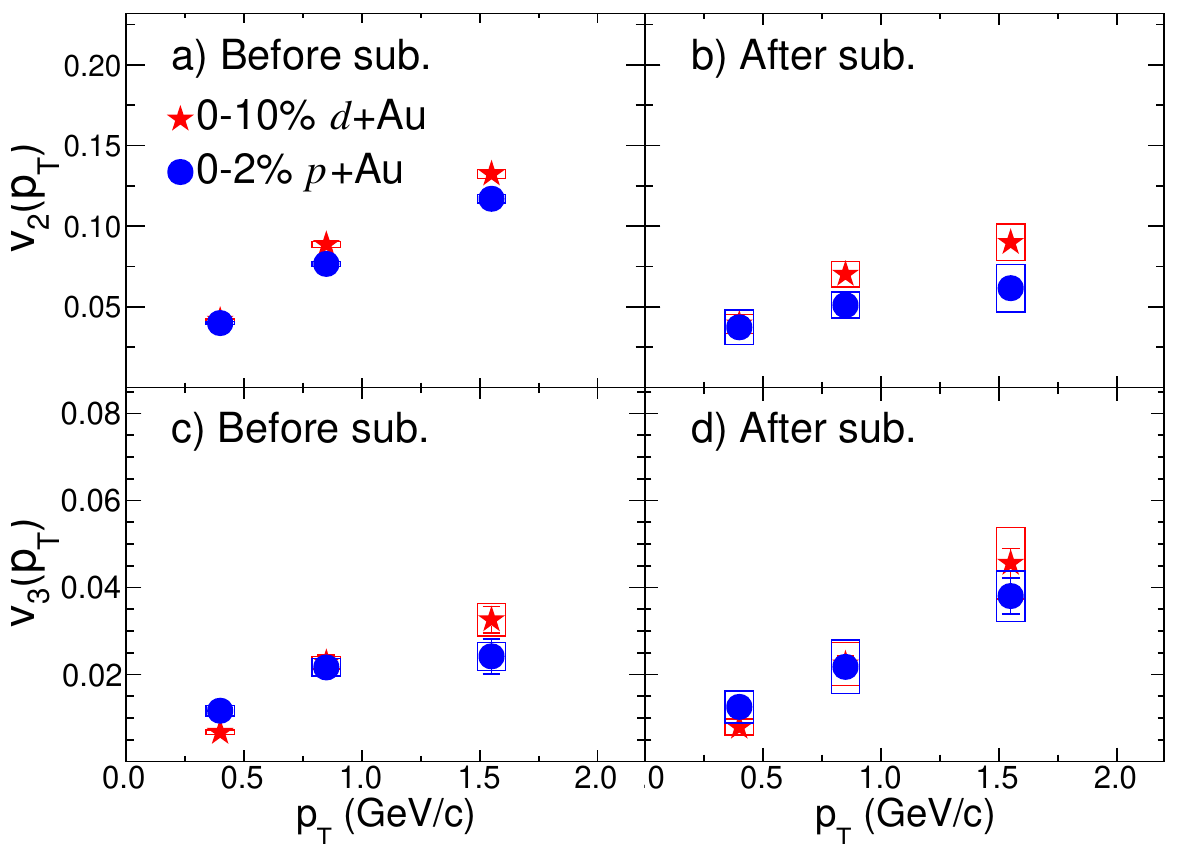}
\caption{Comparison of $v_2(\pT)$ (top row) and $v_3(\pT)$ (bottom row) values based on the $c_1$ method in the 0--2\% most central \pau\ and 0--10\% most central \dau\ collisions before (left column) and after (right column) non-flow subtractions. The boxes are the systematic uncertainties.}\label{fig:17}
\end{figure}

\begin{figure}[ht]
\centering
\includegraphics[width=1.\linewidth]{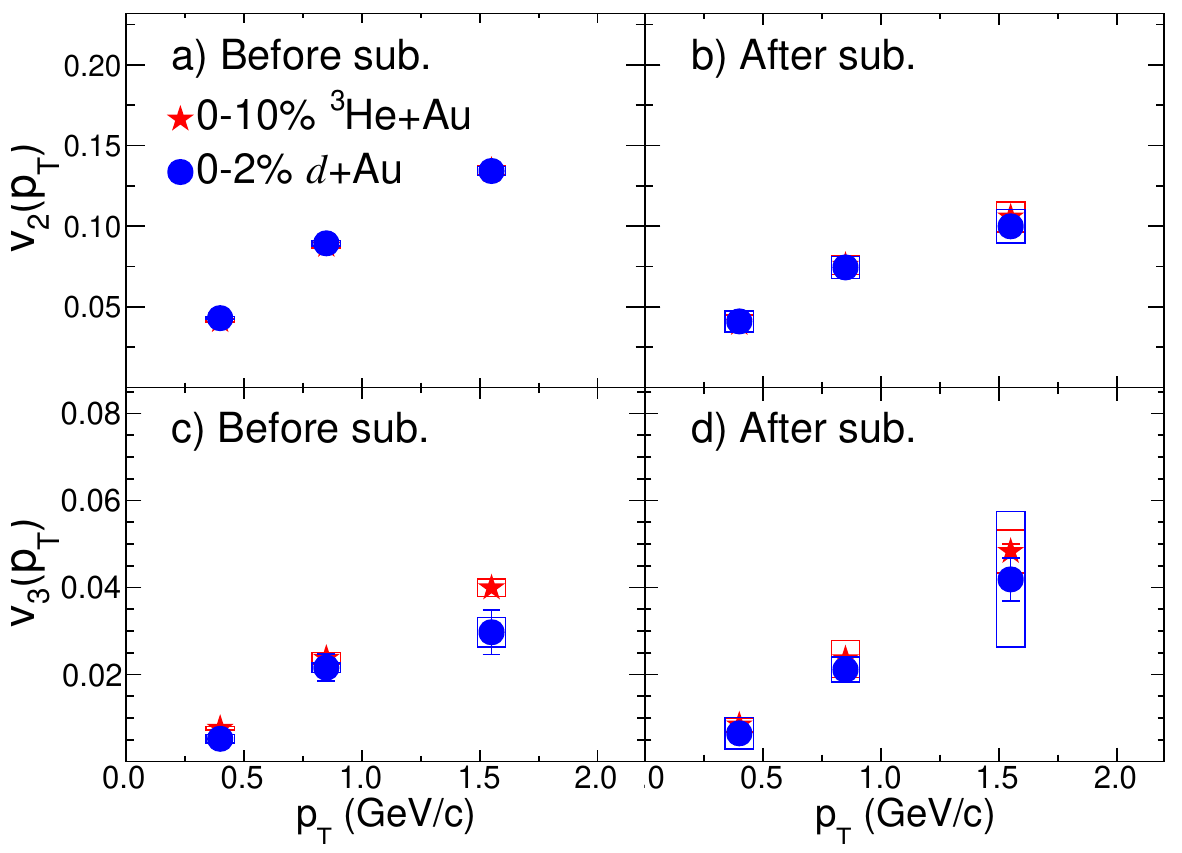}
\caption{Comparison of $v_2(\pT)$ (top row) and $v_3(\pT)$ (bottom row) based on the $c_1$ method in the 0--2\% most central \dau\ and 0--10\% most central \heau\ collisions before (left column) and after (right column) non-flow subtractions. The boxes are the systematic uncertainties.}\label{fig:18}
\end{figure}

\begin{figure*}[htbp]
\includegraphics[width=1\linewidth]{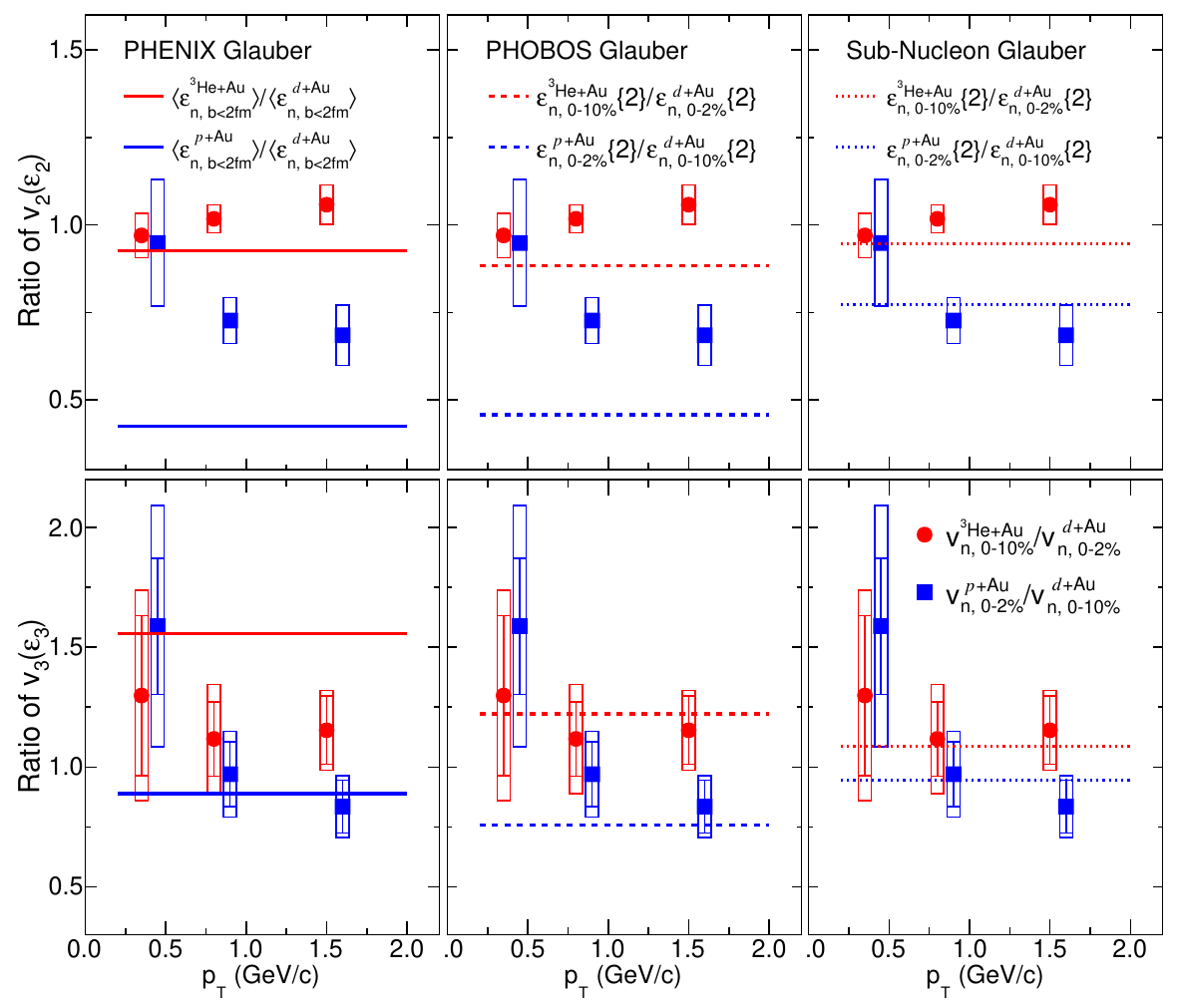}
\caption{ Ratios of $v_2(\pT)$ (top) and $v_3(\pT)$ (bottom) values obtained from the $c_1$ method between the 0--2\% \pau\ and 0--10\% \dau\ collisions (circles) and between the 0--2\% \dau\ and 0--10\% \heau\ collisions (squares). They are compared to ratios of $\varepsilon_{2}$ and $\varepsilon_{3}$ values from three types of Glauber models: PHENIX Glauber model~\cite{PHENIX:2018lia} calculated with simple mean $\lr{\varepsilon_{n}}$ for $b<2$~fm (left column), PHOBOS Glauber model from Ref.~\cite{STAR:2022pfn,PHOBOSGlauber} calculated with $\varepsilon_{n}\{2\}\equiv\sqrt{\lr{\varepsilon_{n}^2}}$ (middle column) and quark Glauber model from Ref.~\cite{Welsh:2016siu} calculated with $\sqrt{\lr{\varepsilon_{n}^2}}$ (right column). The boxes are the systematic uncertainties.}\label{fig:19}
\end{figure*}

\subsection{Comparison of $v_n$ between different systems at similar multiplicity and constraining the initial geometry}\label{sec:5.2}
A notable observation at the LHC is the striking similarity in both the magnitude and $\pT$ dependencies of $v_3$ between $p$+Pb and Pb+Pb collisions at the same multiplicity~\cite{CMS:2016fnw,ATLAS:2014qaj}. This phenomenon has led to the concept of conformal scaling~\cite{Basar:2013hea}, which suggests that the ratios $v_n/\varepsilon_n$ should primarily depend on the charged particle multiplicity density ($dN_{\mathrm{ch}}/d\eta$). The underlying reasoning is that the hydrodynamic response is governed by the ratio of the mean free path to system size, a relationship that follows a power-law function of $dN_{\mathrm{ch}}/d\eta$~\cite{Basar:2013hea}. While this conformal scaling has been validated in large collision systems, such as in comparisons between Au+Au and U+U collisions~\cite{Giacalone:2021udy}, it was also proven effective for $v_2$ in $p$+Pb vs. Pb+Pb~\cite{Basar:2013hea} and $v_3$ in all systems when accounting for possible oversubtraction in \pp\ collisions~\cite{ATLAS:2018ngv}. Given that $v_n$ is primarily driven by final state effects, it is reasonable to expect a similar universal scaling behavior in small systems at RHIC energy.

For two systems, A and B, at similar charged particle multiplicities, we expect the following relation to hold,
\begin{align}\label{eq:comp1}
\frac{v_n^{\rm A}}{v_n^{\rm B}} \approx \frac{\varepsilon_n^{\rm A}}{\varepsilon_n^{\rm B}}\;.
\end{align}
This equation implies that the ratio of $v_n$ between two systems effectively cancels out most of the final state effects, thereby providing a means to constrain the ratio of their eccentricities.

This comparative analysis can be conducted using the centrality selections outlined in Table~\ref{tab:2}. We observe that the average charged particle multiplicities $\lr{N_{\mathrm{ch}}}$ are similar between the 0--2\% \pau\ and 0--10\% \dau\ systems, as well as between the 0--2\% \dau\ and 0--10\% \heau\ systems. Figures~\ref{fig:17} and \ref{fig:18} present these two comparisons, respectively.

The results for $v_2$ and $v_3$ before and after non-flow subtraction show remarkably similar behaviors for 0--2\% \dau\ collisions and 0--10\% centrality \heau\ collisions. In contrast, while $v_3$ values are comparable between 0--2\% \pau\ and 0--10\% \dau\ collisions, there is an approximately 20\% difference in $v_2$.

To further quantify these observations, we calculate the ratios of $v_n$ between 0--2\% \pau\ and 0--10\% \dau, as well as between 0--2\% \dau\ and 0--10\% \heau. These ratios are displayed in Figure~\ref{fig:19}. Systematic uncertainties, including those arising from non-flow subtraction methods, are largely correlated across different systems. The total uncertainties are approximately 5\% for $v_2^{\scriptsize ^3\rm{HeAu}}/v_2^{\scriptsize d\rm{Au}}$ and range from 10\% to 20\% for $v_2^{\scriptsize p\rm{Au}}/v_2^{\scriptsize d\rm{Au}}$. For $v_3$, the uncertainties are larger, particularly at the lowest $\pT$ bin, but decrease to below 20\% at higher $\pT$. The ratio $v_2^{\scriptsize p\rm{Au}}/v_2^{\scriptsize d\rm{Au}}$ is about 20\% below unity, indicating that $\varepsilon_2^{\scriptsize p\rm{Au}}$ is smaller than $\varepsilon_2^{\scriptsize d\rm{Au}}$ by a similar margin. In contrast, the $v_3$ ratios are close to unity, although $v_3^{\scriptsize ^3\rm{HeAu}}$ is systematically larger than $v_3^{\scriptsize d\rm{Au}}$ by roughly 10\%, albeit within sizable uncertainties. This suggests that $\varepsilon_3$ in the three systems at similar multiplicities are roughly comparable.

Next, we compare the ratios of $v_n$ to those of $\varepsilon_n$ from three types of Glauber model calculations. These models include fluctuations at nucleon level~\cite{Nagle:2013lja,PHOBOSGlauber,STAR:2022pfn} or at both nucleon and subnucleon level~\cite{Welsh:2016siu}. The definition of eccentricity also varies depending on whether it is defined as a simple mean $\lr{\varepsilon_n}$~\cite{Nagle:2013lja} or root-mean-square $\varepsilon_{n}\{2\}\equiv\sqrt{\lr{\varepsilon_n^2}}$~\cite{PHOBOSGlauber,STAR:2022pfn,Welsh:2016siu}. The latter definition yields larger values due to the inclusion of event-by-event fluctuations and shows smaller hierarchical differences between the three systems (see Table~\ref{tab:1}). Since $v_n$ measured by the two-particle correlation method is effectively $\sqrt{\lr{v_n^2}}$, i $\varepsilon_{n}\{2\}$ is a more appropriate choice.

Figure~\ref{fig:19} compares the ratios of $\varepsilon_n$ from these three types of Glauber models, calculated for the same centrality range. The two models, PHENIX Glauber~\cite{PHENIX:2018lia,Nagle:2013lja} and PHOBOS Glauber~\cite{STAR:2022pfn,PHOBOSGlauber} without subnucleon fluctuations, fail to reproduce the hierarchy of $v_n$ ratios indicated by the data. These models predict substantially smaller $\varepsilon_2$ values for \pau\ than for \dau\ collisions and a greater $\varepsilon_3$ for \heau\ than for \dau\ collisions, a prediction at odds with the data. However, the model defining eccentricity as its RMS value predicts a smaller difference between \heau\ and \dau.

On the other hand, the Glauber model~\cite{Welsh:2016siu} that includes subnucleon fluctuations yields $\varepsilon_2$ and $\varepsilon_3$ ratios that agree with the data.  It supports the hypothesis $\varepsilon_3^{\scriptsize ^3\rm{HeAu}}/\varepsilon_3^{\scriptsize d\rm{Au}}>\varepsilon_3^{\scriptsize d\rm{Au}}/\varepsilon_3^{\scriptsize p\rm{Au}}\approx 1$, indicating that $\varepsilon_3^{\scriptsize ^3\rm{HeAu}}$ is larger than $\varepsilon_3^{\scriptsize d\rm{Au}}$ by approximately 10\%. 

We concluded that the Glauber model incorporating subnucleonic fluctuations exhibits a hierarchy of $\varepsilon_{n}$ that is consistent with the observed $v_n$ by STAR.
\begin{eqnarray}\label{eq:comp2}
& &\varepsilon_{2}^{\scriptsize ^3\rm{HeAu}}\approx \varepsilon_{2}^{\scriptsize  d\rm{Au}} > \varepsilon_{2}^{\scriptsize p\rm{Au}}\;,\\
& &\varepsilon_{3}^{\scriptsize^3\rm{HeAu}}\approx \varepsilon_{3}^{\scriptsize d\rm{Au}} \approx \varepsilon_{3}^{\scriptsize p\rm{Au}}\;.
\end{eqnarray}

\section{Summary}\label{sec:6}
We presented measurements of elliptic flow ($v_2$) and triangular flow ($v_3$) in high-multiplicity $p$/$d$/$^{3}$He+Au collisions at $\snn$ = 200 GeV. The measurements are performed using two-particle azimuthal angular correlations at mid-rapidity as a function of $\pT$. 

To account for non-flow contributions, which are correlations not associated with collective flow, we estimate these contributions using minimum-bias \pp\ collisions at the same energy and subtract them from the $p$/$d$/$^{3}$He+Au collision data. We employ four non-flow subtraction methods to quantify the uncertainties associated with this process. While non-flow contributions had a significant impact on $v_3$ before subtraction, the $v_3$ values after subtraction are consistent across different pseudorapidity gap selections. We also examined potential biases introduced by selecting high-multiplicity events using alternative criteria, finding overall agreement except for $v_2$ in \pau\ collisions for the $c_0$ subtraction methods.

Additionally, we perform a closure test of the non-flow subtraction procedure using events generated by the HIJING model. The closure level is generally within the quoted systematic uncertainties, with a few exceptions: $v_2$ results might be underestimated (oversubtracted) at high $\pT$, particularly in \pau\ collisions, while $v_3$ results could be slightly overestimated (undersubtracted) by approximately 10\% across all systems and $\pT$ ranges.

Importantly, the systematic uncertainties of $v_n$ largely cancel out when forming ratios of $v_n$ across the three collision systems with comparable charged particle multiplicities. This observation supports a clear ordering of their magnitudes: $v_{2}^{\scriptsize ^3\rm{HeAu}}\approx v_{2}^{\scriptsize d\rm{Au}} > v_{2}^{\scriptsize p\rm{Au}}$, and similarly, $v_{3}^{\scriptsize ^3\rm{HeAu}}\approx v_{3}^{\scriptsize d\rm{Au}} \approx v_{3}^{\scriptsize p\rm{Au}}$. These orderings align with the predictions of eccentricities considering subnucleon fluctuations in the initial geometry.

However, these observed orderings differ from those observed by the PHENIX experiment, which measured correlations between particles at mid-rapidity and particles in the backward rapidity direction on the Au-going side. The PHENIX results are more consistent with an initial geometry that includes only nucleon fluctuations. A state-of-the-art hydrodynamic model analysis~\cite{Zhao:2022ugy} suggests that this discrepancy could, in part, be attributed to longitudinal decorrelations of $v_3$ between mid-rapidity and backward rapidity. Moreover, models that incorporate pre-equilibrium flow but exclude subnucleon fluctuations can also reproduce the measured $v_3$ values. 

In summary, our results highlight the importance of considering subnucleon fluctuations and longitudinal decorrelations in interpreting flow measurements in small collision systems, and they underscore the need for continued refinement of both additional measurements and theoretical models.

\section{Acknowledgement}
We thank the RHIC Operations Group and RCF at BNL, the NERSC Center at LBNL, and the Open Science Grid consortium for providing resources and support.  This work was supported in part by the Office of Nuclear Physics within the U.S. DOE Office of Science, the U.S. National Science Foundation, National Natural Science Foundation of China, Chinese Academy of Science, the Ministry of Science and Technology of China and the Chinese Ministry of Education, the Higher Education Sprout Project by Ministry of Education at NCKU, the National Research Foundation of Korea, Czech Science Foundation and Ministry of Education, Youth and Sports of the Czech Republic, Hungarian National Research, Development and Innovation Office, New National Excellency Programme of the Hungarian Ministry of Human Capacities, Department of Atomic Energy and Department of Science and Technology of the Government of India, the National Science Centre and WUT ID-UB of Poland, the Ministry of Science, Education and Sports of the Republic of Croatia, German Bundesministerium f\"ur Bildung, Wissenschaft, Forschung and Technologie (BMBF), Helmholtz Association, Ministry of Education, Culture, Sports, Science, and Technology (MEXT), Japan Society for the Promotion of Science (JSPS) and Agencia Nacional de Investigaci\'on y Desarrollo (ANID) of Chile.
\bibliography{ref}{}\bibliographystyle{apsrev4-1}

\begin{figure*}[h!]
\includegraphics[width=1.0\linewidth]{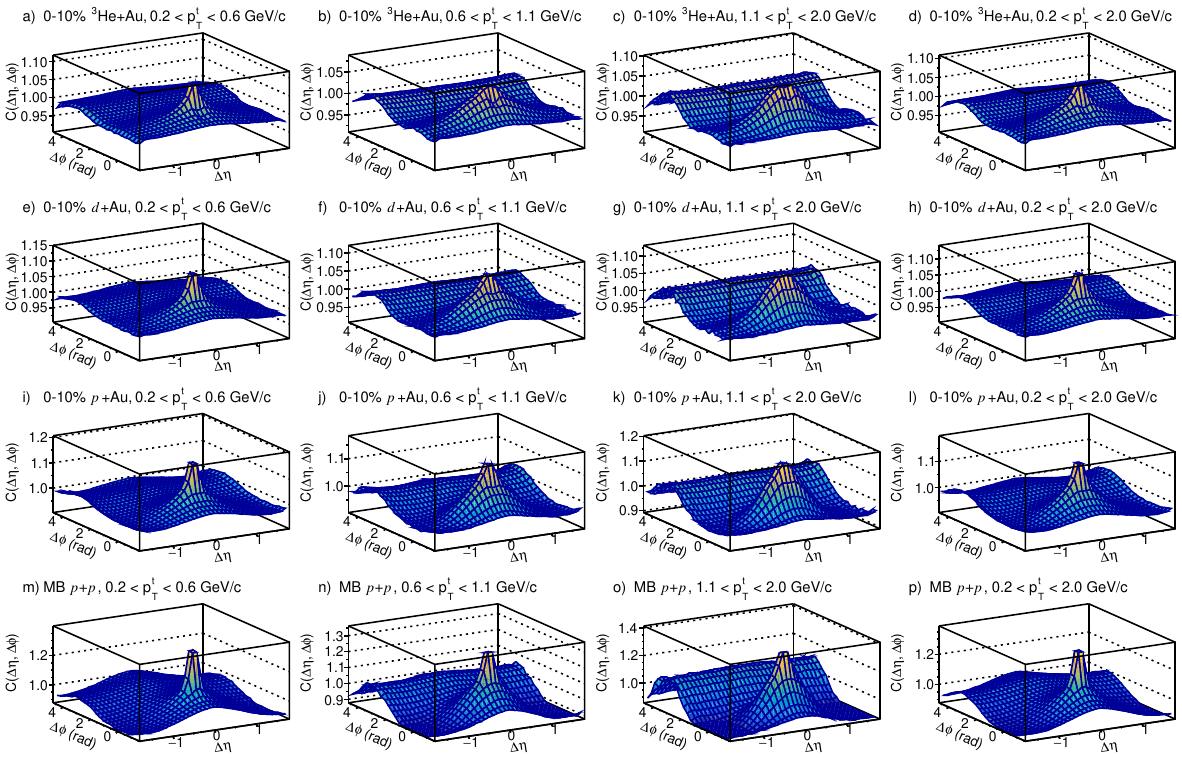}
\caption{The normalized two-particle correlation function as a function of \Deta\ and \Dphi\ for the trigger particles within different $\pT$ ranges (from left to right) in the MB \pp\ and the top 0--10\% \pau, \dau, and \heau\ collisions at \TopE (from bottom to top).}\label{fig:app1}
\end{figure*}

\section{Appendix: Additional plots}\label{sec:app}
\begin{figure*}[htbp]
\includegraphics[width=0.9\linewidth]{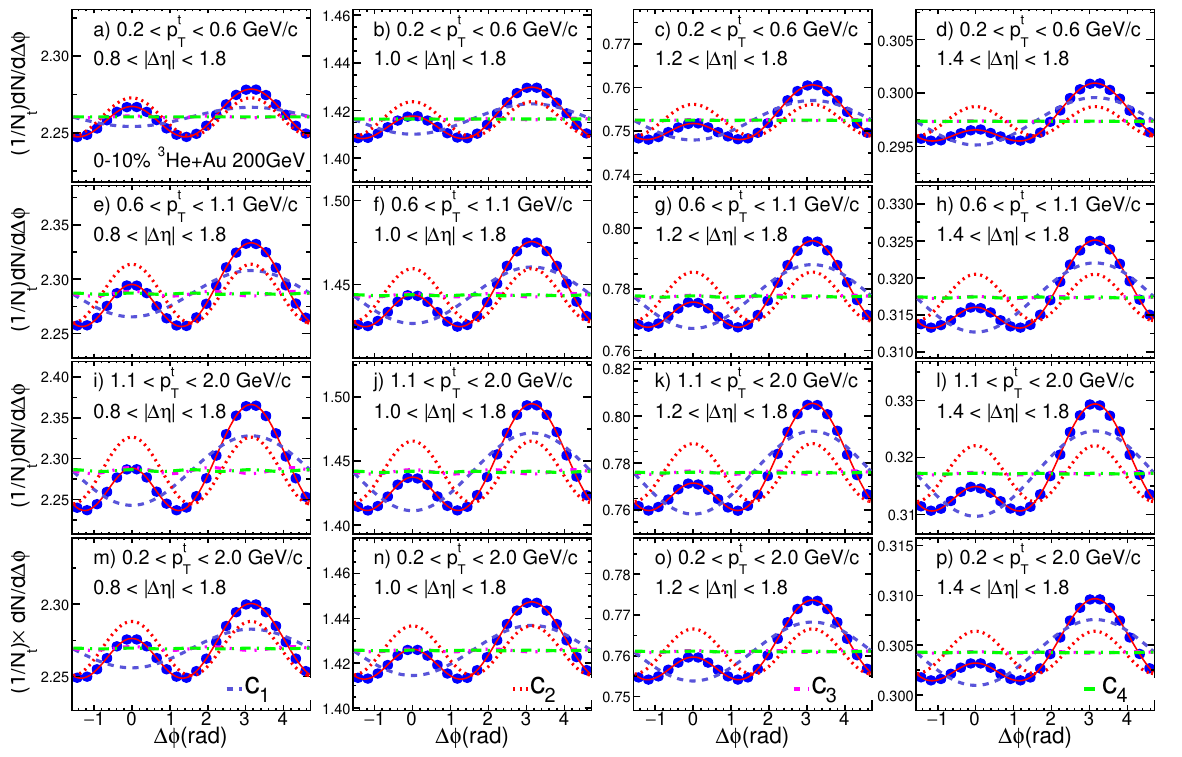}
\caption{The per-trigger yield, \Yphi, as a function of $\dphi$ for trigger particles across different $\pT$ ranges (top to bottom) and different $\Deta$ selections (left to right) in the 0--10\% most central \heau\ collisions at \TopE. The color curves in each panel represent the Fourier components obtained from the Fourier expansion of the per-trigger yield.}
\label{fig:app2}
\end{figure*}

\begin{figure*}[htbp]
\includegraphics[width=0.9\linewidth]{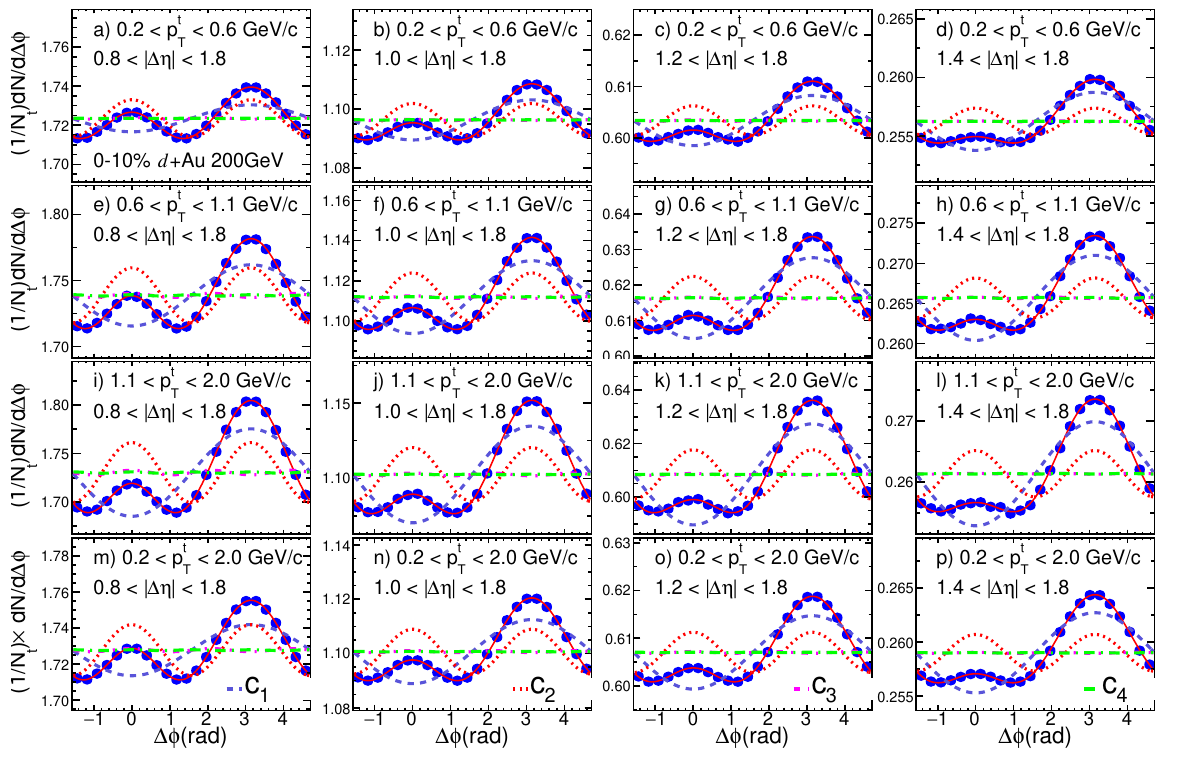}
\caption{The per-trigger yield, \Yphi, as a function of $\dphi$ for trigger particles across different $\pT$ ranges (top to bottom) and different $\Deta$ selections (left to right) in the 0--10\% most central \dau\ collisions at \TopE. The color curves in each panel represent the Fourier components obtained from the Fourier expansion of the per-trigger yield.}
\label{fig:app3}
\end{figure*}

\begin{figure*}[htbp]
\includegraphics[width=0.9\linewidth]{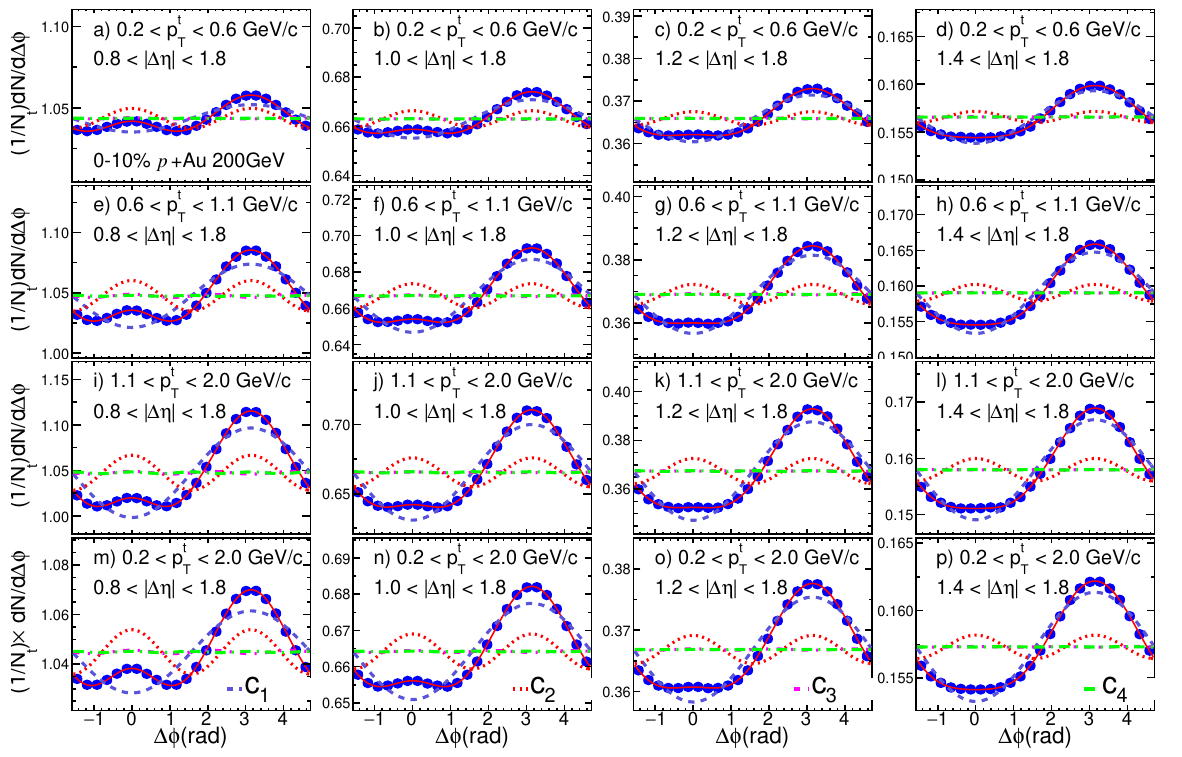}
\caption{The per-trigger yield, \Yphi, as a function of $\dphi$ for trigger particles across different $\pT$ ranges (top to bottom) and different $\Deta$ selections (left to right) in the 0--10\% most central \pau\ collisions at \TopE. The color curves in each panel represent the Fourier components obtained from the Fourier expansion of the per-trigger yield.}
\label{fig:app4}
\end{figure*}

\begin{figure*}[htbp]
\includegraphics[width=0.9\linewidth]{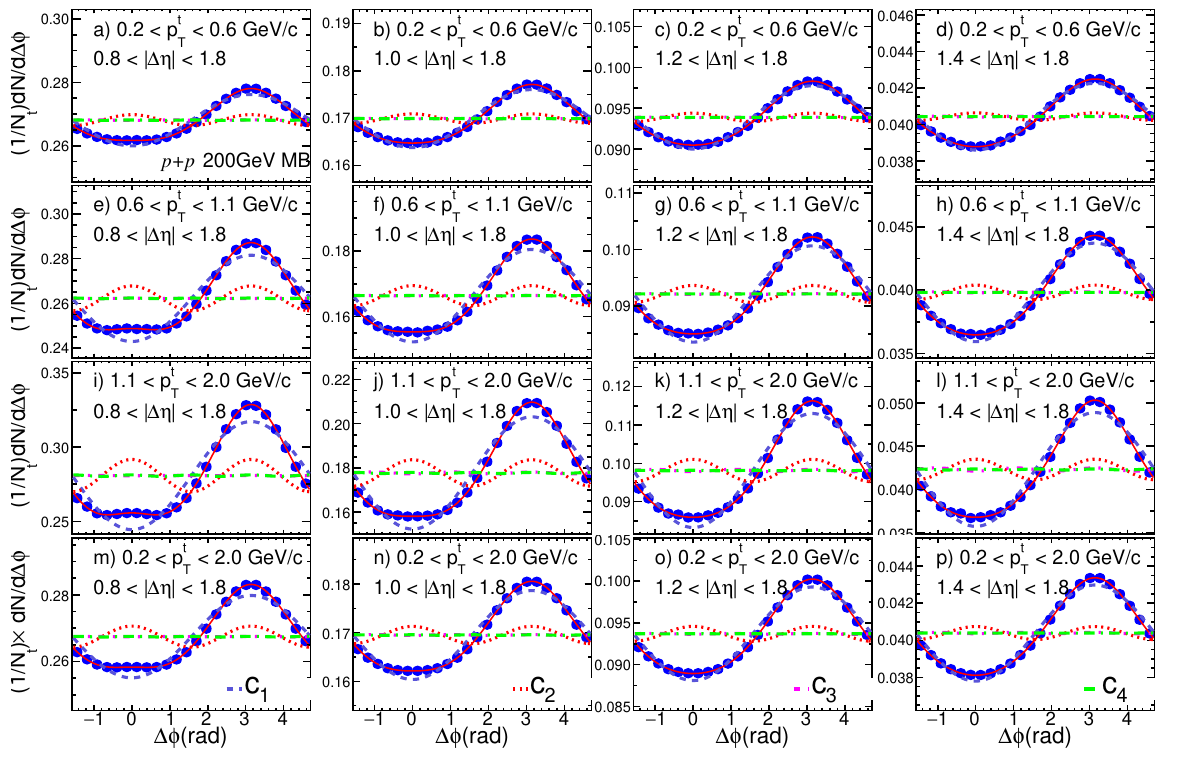}
\caption{The per-trigger yield, \Yphi, as a function of $\dphi$ for trigger particles across different $\pT$ ranges (top to bottom) and different $\Deta$ selections (left to right) in the 0--10\% most central \pp\ collisions at \TopE. The color curves in each panel represent the Fourier components obtained from the Fourier expansion of the per-trigger yield.}
\label{fig:app5}
\end{figure*}

\begin{figure*}[htbp]
\includegraphics[width=0.9\linewidth]{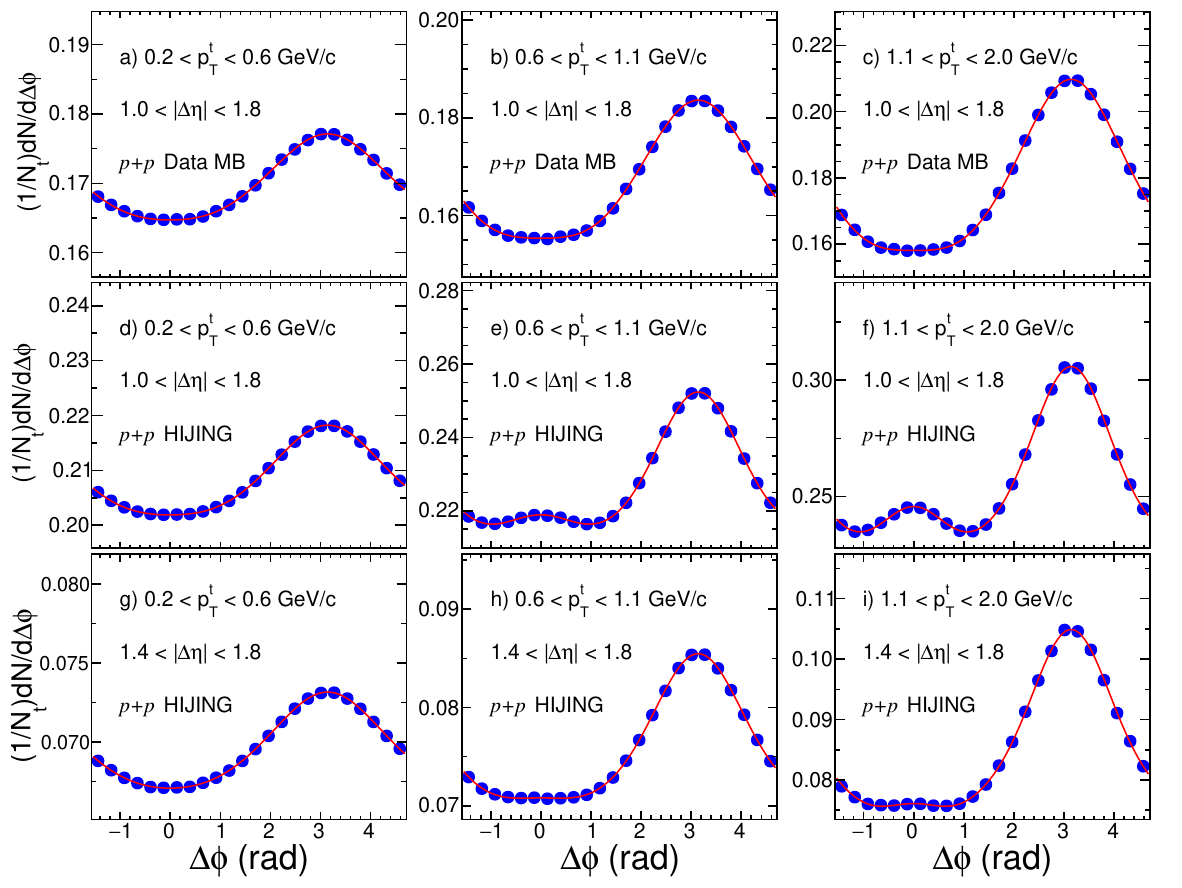}
\caption{The per-trigger yield, \Yphi, as a function of $\dphi$ for trigger particles across different $\pT$ (left to right) and different \Deta\ selections (top to bottom) in \pp\ collisions at \TopE, from both the STAR data (top row) and HIJING event generator (middle and bottom rows). The red lines through the data point represent a fit including the first four Fourier harmonics.}
\label{fig:app6}
\end{figure*}

In this appendix, we present the original correlations that form the basis for deriving the $v_n$ results. Figure~\ref{fig:app1} shows the two-dimensional correlation functions across four $\pT$ ranges from various collision systems. By analyzing these correlation functions, one can extract one-dimensional correlation functions within different $\Deta$ intervals, which are then converted into per-trigger yields. These per-trigger yields are illustrated in Figs.~\ref{fig:app2}-\ref{fig:app5}. To provide additional context, Fig.~\ref{fig:app6} compares per-trigger yields in minimum-bias \pp\ collisions between the experimental data and predictions from the HIJING model.

\begin{figure*}[htbp]
\centering
\includegraphics[width=0.85\linewidth]{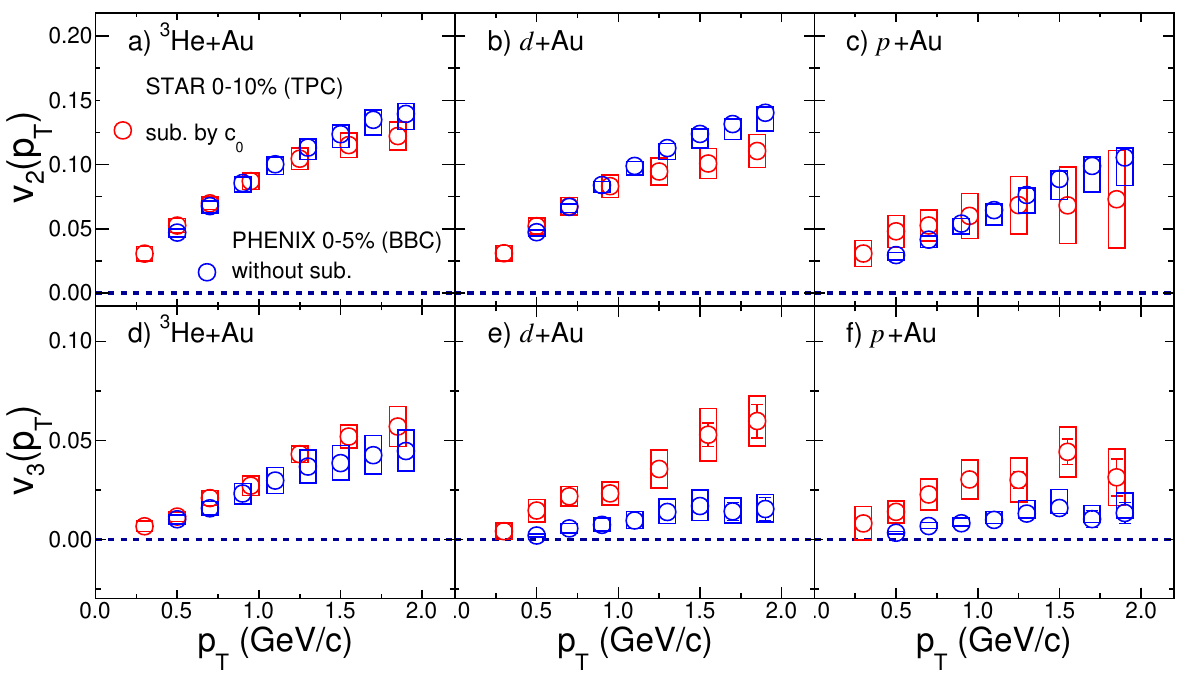}
\caption{Comparison of $v_2(\pT)$ (top row) and $v_3(\pT)$ (bottom row) between measurements obtained by PHENIX (open blue circles) and STAR results with non-flow subtraction based on $c_0$ method (red open circles). The boxes indicate the systematic uncertainties. The PHENIX results are for 0--5\% centrality, while the STAR results are for 0--10\% centrality.}
\label{fig:app7}
\end{figure*}
\end{document}